\RequirePackage[2020-02-02]{latexrelease} % otherwise we get endgroup error WHY?
\documentclass[%
 reprint,superscriptaddress,%amsmath,amssymb,
 aps,pre,shortbibliography
]{revtex4-2}

\usepackage{graphicx}% Include figure files
\usepackage{dcolumn}% Align table columns on decimal point
\usepackage{amsmath,amssymb,bm}% bold math
\usepackage{xcolor}
\newcommand{\blue}[1]{{\textcolor{black}{#1}}}
\begin{document}

%\preprint{APS/123-QED}

\title{Lagrangian Curvature Statistics from Gaussian Subensembles in Turbulent Flows}% Force line breaks with \\
%\thanks{A footnote to the article title}%

\author{Yasmin Hengster}
\affiliation{%
 School of Mathematics and Maxwell Institute for Mathematical Science, University of Edinburgh, UK
}%

\author{Johannes Bosbach}
\author{Daniel Schanz}
\author{Andreas Schr\"oder}
\altaffiliation{
also at Brandenburgische Technische Universit\"at (BTU), Cottbus-Senftenberg, Germany
}
 %\homepage{http://www.Second.institution.edu/~Charlie.Author}
\affiliation{
German Aerospace Center (DLR), Institute of Aerodynamics and Flow Technology, G\"ottingen, Germany
}%
\author{Moritz Linkmann}%
 \email{moritz.linkmann@ed.ac.uk}
\affiliation{%
 School of Mathematics and Maxwell Institute for Mathematical Science, University of Edinburgh, UK
}%

\date{\today}% It is always \today, today,
             %  but any date may be explicitly specified

\begin{abstract}  %<= 600 characters

A salient feature of fully turbulent flows far from onset is the intermittent
occurrence of extreme fluctuations at small spatial and temporal
scales. These have a qualitative and quantitative effect on the instantaneous curvature of a tracer particle trajectory as an intrinsically multi-scale observable.  
Here, we provide a complete %closed? 
statistical description  
of the curvature of tracer particle trajectories in 
%weakly anisotropic
turbulent flows that includes and quantifies intermittency effects. We derive an exact expression and a closed-form
approximation for the curvature probability density function,
both agree well with data obtained from laboratory experiments of different types of turbulent flows, and quantify the generic behavior of the system. The method can be extended to more complex systems such as plasma turbulence.

\end{abstract}

%\keywords{Suggested keywords}%Use showkeys class option if keyword
                              %display desired
\maketitle
% Introduction, see Xu, Wilczek, etc, mention magnetic field line curvature (Yan)

%Reduction of statistical complexity, intermittency

%In efforts to understand the phenomenon of intermittency in turbulent flows, 
%classical approaches such as the Eulerian refined similarity hypothesis 
%\cite{Kolmogorov62,Obukhov62} rely on the 
\section{Introduction}

The dynamic geometry of turbulence, described by the curvature of streamlines, tracer particle trajectories, material loops and, in case of plasma turbulence, magnetic field lines, is connected with several significant effects in such flows. It determines pressure distribution and lift \cite{Anderson2016}, results in %several types of flow
instabilities \cite{DrazinReid2004, PakdelMcKinley1996}, 
and drives magnetic field amplification \cite{Moffatt1978} and particle heating \cite{Drake2006,Dahlin2014} 
%\blue{and } 
in plasma turbulence, to name only a few. Despite its importance, there is no complete %closed
theoretical description of curvature statistics in turbulence, due to the intrinsic statistical complexity and multi-scale nature of such flows. The underlying open problem being intermittency, that is, the absence of statistical self-similarity across
scales, a characteristic feature of strongly turbulent flows that so far eludes theoretical understanding and results in considerable difficulty in the derivation of a fundamental `theory of turbulence' and of high-fidelity turbulence models. 
% universality in "geometry of turbulence"
% statistical description of geometric properties of turbulence
% Grauer -- synthetic turbulence -> superposition of gaussian processes 
%
%These difficulties have so far also precluded a statistical description of the geometric properties of turbulence. 
%
%The trajectory of a tracer particle that samples regions of the flow with different turbulence intensity is a three-dimensional space curve, and as such can be described completely by its instantaneous curvature $\kappa$ and its torsion \cite{Braun2006}.  In the Frenet-Serret frame, $\kappa = a_n/u^2$, where $a_n$ is the absolute value of the acceleration component normal to the velocity, and $u$ the velocity magnitude. As such, the curvature statistics depend on large and small-scale statistics in a turbulent flow, and intermittency should have a quantifiable effect on it. At present this is neither known nor understood.

% brief summary of results for HIT, vK experiments and rotating RBC 
Curvature probability density functions (PDFs) have been measured in variety of turbulent flows, in homogeneous and
isotropic turbulence (HIT) using data obtained by direct numerical simulation (DNS)
\cite{Braun2006,Scagliarini2011}, and in the laboratory for von K\'arm\'an flow (vKF) \cite{Xu2007,
Hengster2023}, rotating \cite{Alards2017} and non-rotating Rayleigh-B\'enard
convection (RBC) \cite{Alards2017, Hengster2023} and in a zero-pressure-gradient
boundary layer over a flat plate \cite{Hengster2023}. In plasma turbulence, according to numerical and observational data, the curvature of 
magnetic field lines is distributed similarly to that of tracer particle trajectories in vKF, RBC and HIT
%: Yang et al – DNS of MHD turbulence and PIC for plasma turbulence - same pdfs for MHD and plasma turbulence, (Bandyopadhyay et al. 2020) MMS, Huang 2020 MMS data [claim agreement with Yang et al, but measure κ0.33 and κ−2.16 , then fit to double Pareto lognormal ] Yong Ji et al MMS data, see intro and figure 4: κ0.8 and κ−2.1
\cite{Schekochihin2001, Schekochihin2002, Schekochihin2004, Yang2019, Bandyopadhyay2020, HuangEA20-Bcurve, Ji2022, Hengster2024Thesis}, and so is the curvature of material loops in isotropic turbulence \cite{Bentkamp2022}. That is, except for the boundary layer, where a strong
unidirectional flow suppresses high-curvature events, curvature PDFs appear to have a universal form \cite{Hengster2023, Hengster2024Thesis}. 

% what we do and summary of of our results
%Statistical reduction techniques
%such as decompositions into subensembles with simpler statistics, 
%where theoretical progress is tangibly possible, can be useful tools in this context. 
%
% turbulent flows decomposed in subensembles according to level of turbulence
% -- each subensemble Gaussian, independent RVs
% -- what can be done with decomposition -> Bentkamp et al
% -- novelties here: moderate level of anisotropy, master curve, full description of curvature pdf
% -- no intermittency in subensembles --> model works in full ensemble due to mapping to single ensemble, rescaling
% -> focus should be more on intermittency modelling, if we remove intermittency and still keep considerable 
%    flow complexity, stats are the same except for variances -- scaling argument
% "Deviations from the peak of the model PDF, however, are
% an indication of fully developed turbulence and may be due
% both to the tails of the acceleration PDF and to the corre-
% lation of acceleration and velocity." -> Xu et al
% peak is unchanged by filtering of kappa -> Xu et al

%TODO: cite Bentkamp material lines nature paper
Here, we \blue{combine a statistical reduction method recently developed by Bentkamp {\em et al.} \cite{Bentkamp2019}, which decomposes the full turbulent ensemble in the Lagrangian frame of reference in approximately Gaussian subensembles, with a model describing the curvature probability density function (PDF) based on Gaussian statistics \cite{Xu2007}. Doing so, we provide a superstatistical  description of, and an approximate closed-form model expression for,} the statistics of the instantaneous curvature of tracer particle trajectories in turbulence, including intermittency effects. 
The derived expressions are compared against 
PDFs 
measured in experiments of turbulent vKF 
and RBC 
\cite{Godbersen2021}, 
showing that they capture both the core and tails of the curvature PDFs across distinct turbulent flows.

The trajectory of a tracer particle that samples regions of the flow with different turbulence intensity is a three-dimensional space curve, and as such can be described completely by its instantaneous
curvature $\kappa$ and its torsion \cite{Braun2006}.  In the Frenet-Serret
frame, $\kappa = a_n/u^2$, where $a_n$ is the absolute value of the
acceleration component normal to the velocity, and $u$ the velocity magnitude. As such, the curvature statistics depend on large and small-scale statistics
in a turbulent flow, and intermittency should have a quantifiable effect on it. At present this is neither known nor understood.
Xu {\em et al.} \cite{Xu2007} derived a closed-form expression of the curvature, assuming velocity and acceleration to be independent Gaussian
vectors. % also in thin layers see Ouellette_07...arxiv (Ouellette & Gollub) -- spatiotemporal chaos, not turbulence
% bfield curvature
%The same arguments have been applied to derive PDFs for magnetic field-line curvature in plasma
%turbulence \cite{Yang2019}. 
%That is, 
Despite the assumptions underlying the derivation of the model PDF to be 
generally not valid for turbulent flows -- acceleration statistics are distinctively non-Gaussian \cite{LaPorta2001,Voth2002,Mordant2004a,Homann2011} 
and Lagrangian velocity and acceleration fluctuations are correlated \cite{Crawford2005} --
the model supplies a 
correct qualitative description of the tails of the curvature PDFs. However, the core of 
the PDF, that is the generic behavior of the system, is not well approximated, %by the Gaussian model, 
most likely because of non-Gaussian statistics and intermittency effects \cite{Xu2007}.
%Because its underlying assumptions are violated by non-Gaussian acceleration statistics and intermittency, existing Gaussian models fail to capture the generic behavior of curvature fluctuations in turbulence. -- You can still mention tail agreement later, where nuance is safer.
To include these effects, we
decompose the full turbulent ensemble in the Lagrangian frame of reference in approximately 
Gaussian subensembles using the statistical reduction method of Bentkamp {\em et al.} \cite{Bentkamp2019}, and consider the model in each subensemble where its underlying assumptions are satisfied to a good approximation.  
%we supply the first complete theoretical description of curvature statistics in weakly anisotropic turbulent flows including effects of spatio-temporal intermittency.

%To supply a complete theoretical description of curvature statistics in 
%weakly anisotropic turbulent flows, we derive an expression for the curvature probability density function
%(PDF) for the ensemble of tracer particle trajectories in turbulence including
%spatio-temporal intermittency. To do so, the turbulent ensemble is decomposed into  
%Gaussian subensembles \cite{Bentkamp2019}, each of which is then shown to satisfy the 
%assumptions of the aformentioned model proposed by Xu {\em et al.} \cite{Xu2007}.
%We obtain a master curve for the PDF for
%near-Gaussian subensembles, obtained by conditioning on the squared
%acceleration coarse-grained over a few viscous time units \cite{Bentkamp2019},
%where an analytic form of the PDF is known \cite{Xu2007}, from which we calculate
%the PDF for the full ensemble.  

%Theory
\section{Theory}
To reduce the statistical complexity of turbulence, 
%in the Lagrangian frame of reference, 
we consider the {\em coarse-grained acceleration} \cite{Bentkamp2019}
\begin{equation}
    \label{eq:alpha}
    \alpha(t) = \int_{-\infty}^{\infty} d\tau \ G_{\Theta}(\tau) |\bm{a}(t+\tau)|^2 
\end{equation}
% better use more compact definition for convolution
where $G_{\Theta}(\tau)$ is a Gaussian filter kernel with standard deviation $\Theta$ and 
$\bm{a}$ the acceleration along a tracer particle trajectory, as a statistical conditioning 
parameter with respect to which the full turbulent ensemble can be decomposed into simpler, approximately Gaussian, 
subensembles. This is akin to approaches in the Eulerian frame of reference where 
near-Gaussian statistics can be obtained by conditioning 
on the dissipation or energy transfer rates averaged over a given 
length scale \cite{Gagne1994,Naert1998, Homann2011,Lawson2019}, %check reference 
inspired by the Eulerian refined similarity hypothesis \cite{Kolmogorov62,Oboukhov62}.
\blue{Within the statistical approach to turbulence, we formally consider the coarse-grained acceleration as a random observable $A$, that inherits its randomness from the formal interpretation of velocity and acceleration being random functions.}

This formalism allows to calculate velocity, velocity increment and acceleration statistics semi-analytically from simpler, here near-Gaussian, PDFs in the subensembles by the law of total probability,
%\begin{equation}
%    p(Y) = \int d\alpha f(\alpha) p_{\alpha}(Y) \ ,
%	\label{eq:reconstruction}
%\end{equation}
%where $Y$ stands for velocity, velocity increments or acceleration, $f(\alpha)$ is the PDF of the coarse-grained acceleration and 
%$p_{\alpha}(Y)$ are the PDFs in the subensembles,
provided their variances and the PDF of $\alpha$ \blue{, $f(\alpha):=P_A(A=\alpha)$ are}  known \cite{Bentkamp2019}. %\yasmin{why do we need the variances here?} ML: a Gaussian is completely described by its variance, and this "parameter" is needed to specify the distribution. 
For homogeneous and isotropic
turbulence, the so reconstructed velocity, velocity increment and acceleration PDFs agree very well with DNS data \cite{Bentkamp2019}, provided 
the filter width $\Theta$ is appropriately chosen, that is, $2\tau_\eta \leqslant \Theta \leqslant 3\tau_\eta$. 
%discuss scaling with alpha and that differences in scaling over increments in $\tau$ are a hallmark of intermittency?
%instrumental in our results, expect more differences in scaling between large and small scales with increasing Re -- our results are consistent with this.
%Here, we validate this formalism for laboratory flows and apply it to derive a complete statistical description for the instantaneous curvature statistics in weakly anisotropic turbulence.

A decomposition in Gaussian subsensembles is particularly useful for the calculation of curvature statistics, as 
the model proposed by Xu {\em et al.} \cite{Xu2007} would be applicable to each Gaussian subsensemble, provided acceleration 
and velocity fluctuations in each subensemble are statistically independent. Under these assumptions, \blue{there is a family of curvature PDFs indexed by $\alpha$, with each single member thereof describing the} curvature PDF \blue{in a particular Gaussian subensemble} 
\begin{align}
p(x_\alpha) = \frac{e^{x_\alpha^{-2}/16}}{16\sqrt{2\pi}x_\alpha^4}
	& \left( \left( 3 + \frac{1}{2x_\alpha^2} \right)K_{3/4}\left( \frac{1}{16 x_\alpha^2} \right) \right . \nonumber \\
      - & \left . \left( 5 + \frac{1}{2x_\alpha^2} \right)K_{1/4}\left( \frac{1}{16 x_\alpha^2} \right)
\right) \ ,
\label{eq:non-dim-pdf}
\end{align}
 where $K_{1/4}$ and $K_{3/4}$ are modified Bessel functions of the second kind \cite{Xu2007} and $x_\alpha:=   \kappa \left(\sigma^{(\alpha)}_u \right)^2 / {\sigma^{(\alpha)}_a}$ is the curvature in each subensemble, non-dimensionalised by the standard deviations $\sigma^{(\alpha)}_u$ and $\sigma^{(\alpha)}_a$ of velocity and acceleration, respectively, in the subensemble. %subensemble corresponding to a particular value of $\alpha$.
%
%The full curvature PDF can then be calculated according to 
Using the law of total probability, the full curvature PDF follows exactly from the Gaussian subensemble statistics
\begin{equation}
    p_\kappa(\kappa) = \int d\alpha f(\alpha) \, p(x_\alpha) \, \sigma^{(\alpha)}_a / \left(\sigma^{(\alpha)}_u \right)^2 
    %p_{\alpha}(\kappa) \ ,
	\label{eq:reconstruction}
\end{equation}
where \blue{$f(\alpha):=P_A(A=\alpha)$} is the PDF of the coarse-grained acceleration,
%and $p_{\alpha}(\kappa) = 
and \blue{the} $p(x_\alpha) \sigma^{(\alpha)}_a / \left(\sigma^{(\alpha)}_u \right)^2  $ are the re-dimensionalised curvature PDFs in the subensembles.

%Data and Methods
% describe experimental configurations, fitting and calibration
\section{Data and Methods}
\label{sec:methods}
We compare our theoretical results against two datasets, turbulent vKF
at $Re_{\lambda} = 270$ and RBC at $Ra = 1.53
\times 10^9$ with $Re_{\lambda} = 183$. The vK experiment took place
at the von K\'arm\'an facility at the MPI G\"ottingen \cite{Schroeder2022}. Two
counter-rotating propellers with $500\, mm$ diameter, installed in
$z$-direction, stir water resulting in approximately homogeneous and
isotropic turbulence in a small volume $V$ in the center of the flow chamber.
The propeller motion also induces large-scale flow, resulting in 
intermittent
%to be intermittent 
large-scales fluctuations and weak anisotropy \cite{Voth2002}.
The flow is seeded with Dynoseeds TS20 tracer particles, illuminated
with high-repetition speed lasers and recorded by four high-frequency cameras to obtain
high temporal resolution. A detailed description of the experiment
can be found in Ref. \cite{Schroeder2022}.

The RBC experiment took place at DLR G{\"o}ttingen \cite{Bosbach2021}, \blue{using a}
cylindrical convection cell of aspect ratio $\Gamma = 1$ and height $H=1.1\, m$, \blue{with air as the working fluid.} 
%The $z$-direction is oriented normal to the heated bottom plate, which is made out of aluminium and electrically heated using temperature sensors and a controller 
%to ensure a constant temperature to ensure a \blue{uniform and steady} constant\blue{-temperature} bottom heating. Constant\blue{-temperature} cooling at the top is achieved by perfusing the top plate with temperature-controlled water from a cooling bath. 
\blue{The bottom plate is made out of aluminium and electrically heated using embedded temperature sensors in combination with a controller to ensure a uniform and steady temperature. Constant-temperature cooling at the upper boundary is achieved by perfusing the top plate with temperature-controlled water from a cooling bath.}
Helium-filled soap bubbles (diameter $< \eta$) with an average life
expectancy of $\sim 330\, s$ are used as tracers \blue{ and illuminated by an array of pulsed LEDs placed above the transparent top plate. The particle images are recorded by a system of six scientific CMOS cameras}. 
%Pulsed LEDs are placed above the top plate to illuminate the particles, and images are recorded by a system of six scientific CMOS cameras. 
%Small-scale turbulent structures and the characteristic large scale circulation (LSC) with its typical dynamics can be observed in this data, see Ref.~\cite{Godbersen2021}. 
\blue{Apart from a variety of small-scale turbulent structures, the characteristic large-scale circulation (LSC) with its typical dynamics can be observed in this data, see Ref.~\cite{Godbersen2021}. Further, a variety of statistical Lagrangian properties of the flow are accessible through this data set \cite{Hengster2023, Weiss2024}.}
Details of the experiment and the set-up are provided in Refs.~\cite{Bosbach2021, Godbersen2021}. 
\blue{In what follows, the $z$-direction is oriented normal to the heated bottom plate.}

%From Daniel's comment
Particle trajectories in both experiments are reconstructed using the Shake-The-Box (STB) algorithm \cite{Schanz2016, Schroeder2023} which  can operate in conditions with a high density of particle images. The algorithm applies Iterative Particle Reconstruction \cite{Wieneke2013, Jahn2021} and a temporal predictor-corrector scheme, which allows tracking a large number of particles without generating ghost tracks. 
%For RBC, up to 500,000 particles could be tracked per time step. % do we need this?
The raw particle tracks are regularized - yielding velocity and acceleration - by applying the {\em TrackFit} algorithm \cite{Gesemann2016}, which fits B-splines of order three to the particle positions with a calibrated filter strength. The experimental error $\Delta x$ on the particle positions is estimated by assuming that the third derivative is noise \cite{Gesemann2016}. Further details on postprocessing and calibration are provided in Refs.~\cite{Gesemann2016, Bosbach2021, Godbersen2021, Hengster2023}. For RBC, we only consider statistics in the bulk. % and disregard data pertaining to thermal and side boundary layers in the analysis. 
Parameters and key observables of both experiments are summarised in table~\ref{tab:flow-properties}.

%Particle positions in both experiments are recorded using the Shake-The-Box
%(STB) algorithm \cite{Schanz2016, Schroeder2023} which can track up to 100000
%particles per time step in a flow with a high density of particles. The
%algorithm applies Iterative Particle Reconstruction \cite{Wieneke2013,
%Jahn2021} and a temporal predictor-corrector scheme, which leads to tracking a
%large number of particles without generating ghost tracks. 
%
%Trajectories are calculated by fitting B-splines of order three using the {\em TrackFit} algorithm \cite{Gesemann2016}, 
%with the experimental error $\Delta x$ on the particle positions estimated by assuming that
%the third derivative is noise \cite{Gesemann2016}.
%Further details on postprocessing and calibration can be found in 
%Ref.~\cite{Gesemann2016, Bosbach2021, Godbersen2021, Hengster2023}.
%We disregard data pertaining to thermal and side boundary layers in the analysis of the RBC data
%and thus only consider statistics in the bulk. 
%Parameters and key observables of
%both experiments are summarised in table \ref{tab:flow-properties}.

\begin{table*}[th]
   \begin{ruledtabular}
    \centering
    \begin{tabular}{cccccccccccc}  
       flow type & Ra & $Re_{\lambda}$ &$\tau_{\eta} [s]$ & $f [Hz]$ & $\tau_{\eta} \cdot f $ & $\eta [mm]$ & $\Delta x [\mu m]$ & $\eta/ \Delta x$ &$f_p [Hz]$ & $V [cm^3]$ & $N$ \\
        \hline
        vKF  & --                 & $270$ & $0.013$ & $1250 $ & 16.25 & $0.1$ & 3  & 33   & $0.5$ & $0.4 \times 0.4 \times 0.15 $ & 92,739,514 \\
        RBC & $1.53 \times 10^9$ & $183$ & $0.18$  & $30 $   & 5.4   & $1.7$ & 50 & 34 &  --    & $ 55^2 \pi \times 110$    & 12,150,782 
    \end{tabular}
    \caption{
	    Experimental parameters and properties of the analysed vKF and RBC)
	    with Rayleigh number Ra, %indicating the level of thermal driving,
	    Taylor-scale Reynolds number $Re_{\lambda}$, $\tau_{\eta}$ and
	    $\eta$ the Kolmogorov time-scale and Kolmogorov length-scale,
	    respectively. The camera frequency is $f$ and the experimental
	    error $\Delta x$ is estimated by spectral analysis
	    \cite{Gesemann2016}. 
     %The effective filter scale introduced by the fitting is denoted by $\ell$. 
     The propeller frequency for vKF 
	    is $f_p$, and the camera system records
	    $N$ trajectories in the volume $V$. For RBC, we exclude thermal
	    and side boundary layers from the analysis, resulting in a smaller 
	    analysis volume of size $ 52.5^2 \pi \times 104.5 \ {\rm cm}^3$.
    }
    \label{tab:flow-properties}
     \end{ruledtabular}
\end{table*}

%Results
\section{Decomposition into approximately Gaussian subensembles}
\label{sec:decomposition}
\begin{figure}
	\centering
    \vspace{-1em}
    \includegraphics[width = \columnwidth]{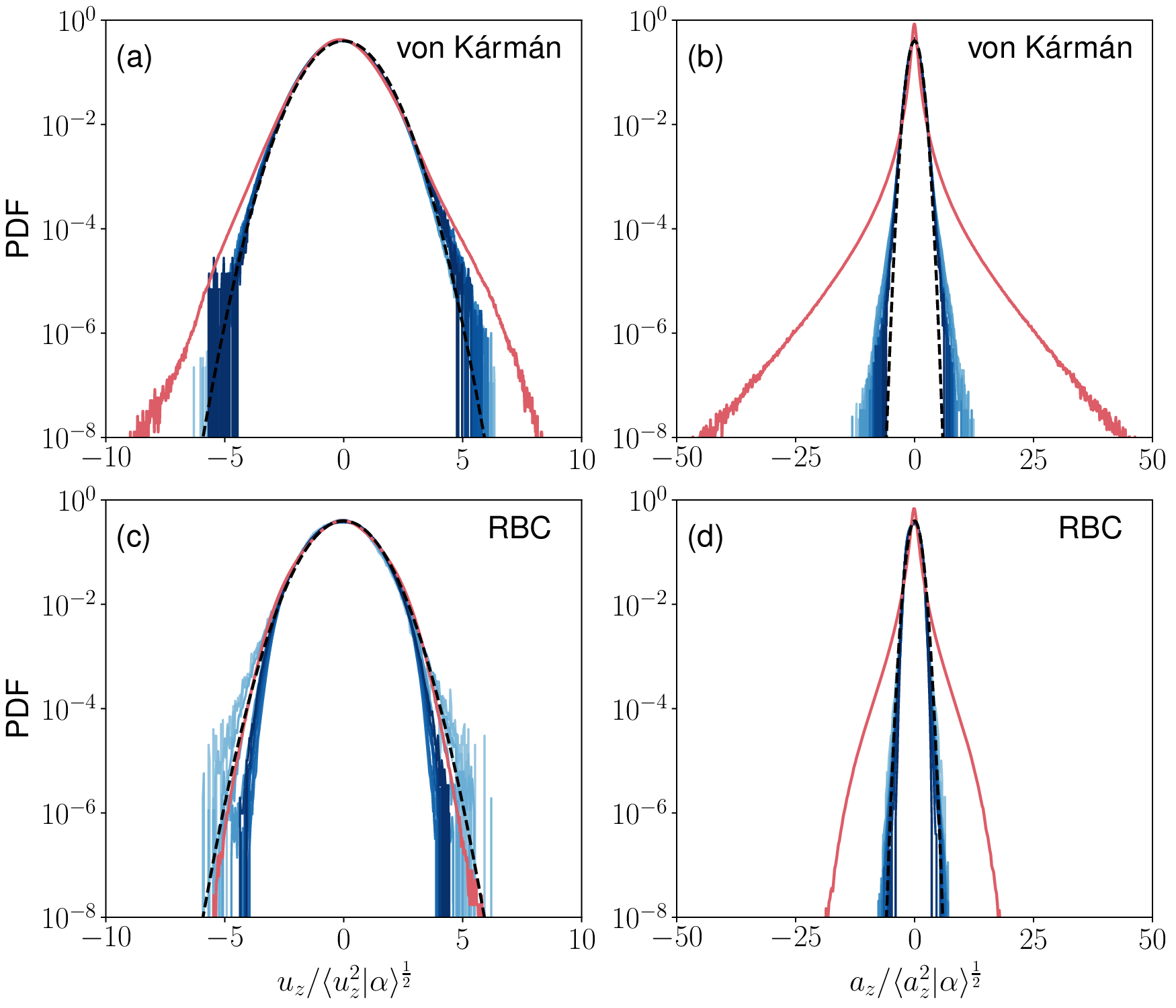}%{z-component.png}
    \vspace{-0.5em}
	\caption{PDFs of the $z$-components of velocity (a,c) and acceleration (b,d) for the full ensemble (red) and in each subensemble (blue) vor vKF and RBC. 
	         The color gradient indicates increasing values of $\alpha$ as darker shades, the dashed line shows 
		 a Gaussian with zero mean and unity standard deviation. 
	         %(a) vKF, velocity, (b) vKF, acceleration, 
	         %(c) RBC, velocity, (b) RBC, acceleration. 
		 %PDFs of $x$- and $y$-components are similar (not shown). %\cite{SM}
   }
    \label{fig:pdfs-subens}
\end{figure}

% Gaussian subensembles
Subensembles for both datasets are generated by conditioning on $\alpha$, obtained according to Eq.~\eqref{eq:alpha} with a filter width of $3 \tau_\eta$. 
As shown in Fig.~\ref{fig:pdfs-subens}(a,b) for vKF and (c,d) for 
RBC, both velocity and acceleration components 
are near-Gaussian distributed in each subensemble. 
Conditional standard PDFs are 
shown in blue with the color gradient indicating increasing values 
of $\alpha$. The red curve pertains to the full ensemble and the dashed
line is a Gaussian with zero mean and unity standard deviation. 
\blue{As all components fluctuate similarly in all subensembles, we only show the PDFs of the $z$-component as representative for the subensembles. However, we observe pronounced non-Gaussian $u_z$-fluctuations for vKF, reflecting the large-scale intermittency and anisotropy present in vKF. The movement of the rotors induces large-scale pumping and shearing modes, that enhance velocity fluctuations in axial direction (pumping mode) and create a shear layer in the centre of the domain due to fluid moving with the propeller direction (shearing mode)  \cite{Voth2002}.}   

\begin{figure*}
	\centering
     %von K\'arm\'an flow \\
    \includegraphics[width =0.85\textwidth]{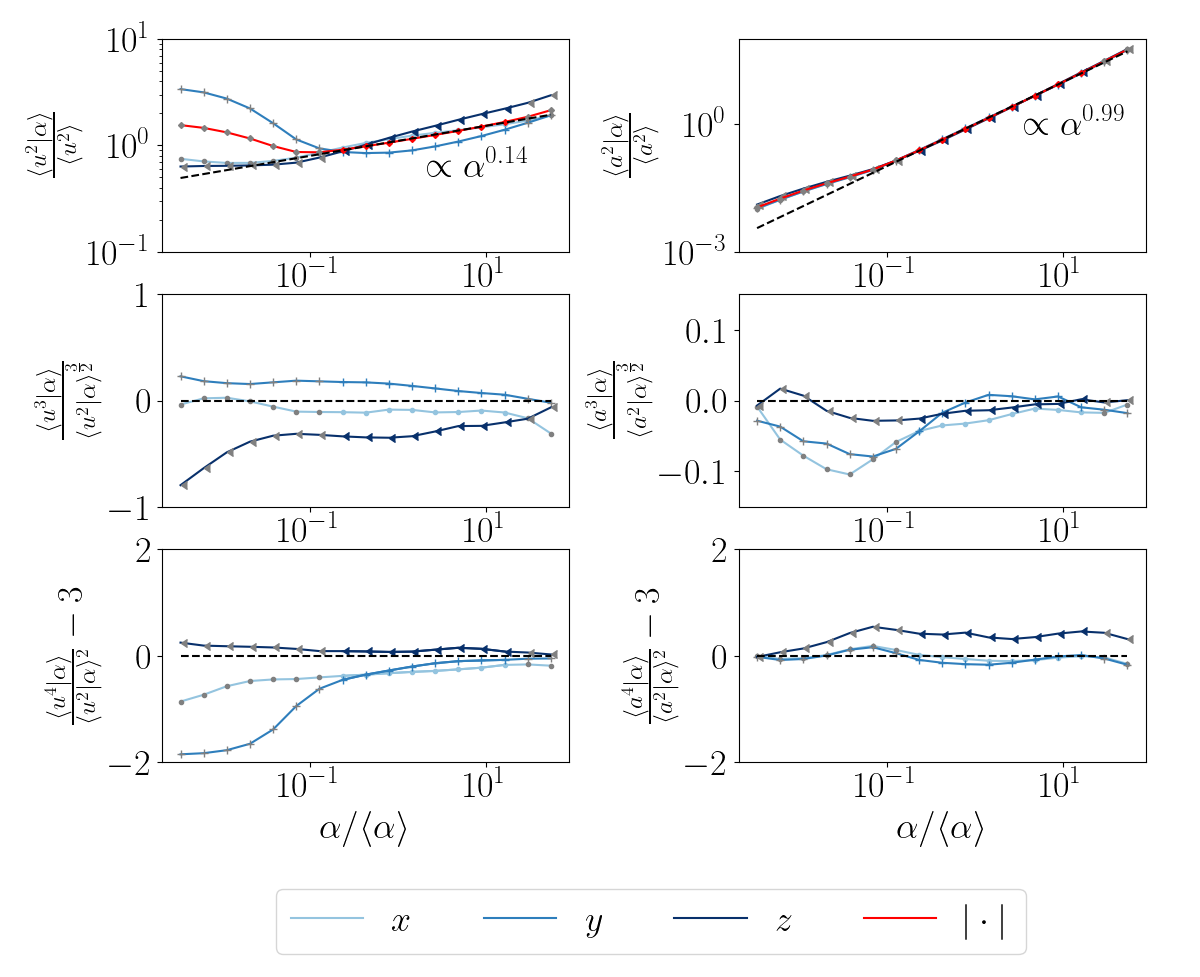} \\
    \vspace{-6em}
    von K\'arm\'an flow \\
    \vspace{1em}
	\includegraphics[width = 0.85\textwidth]{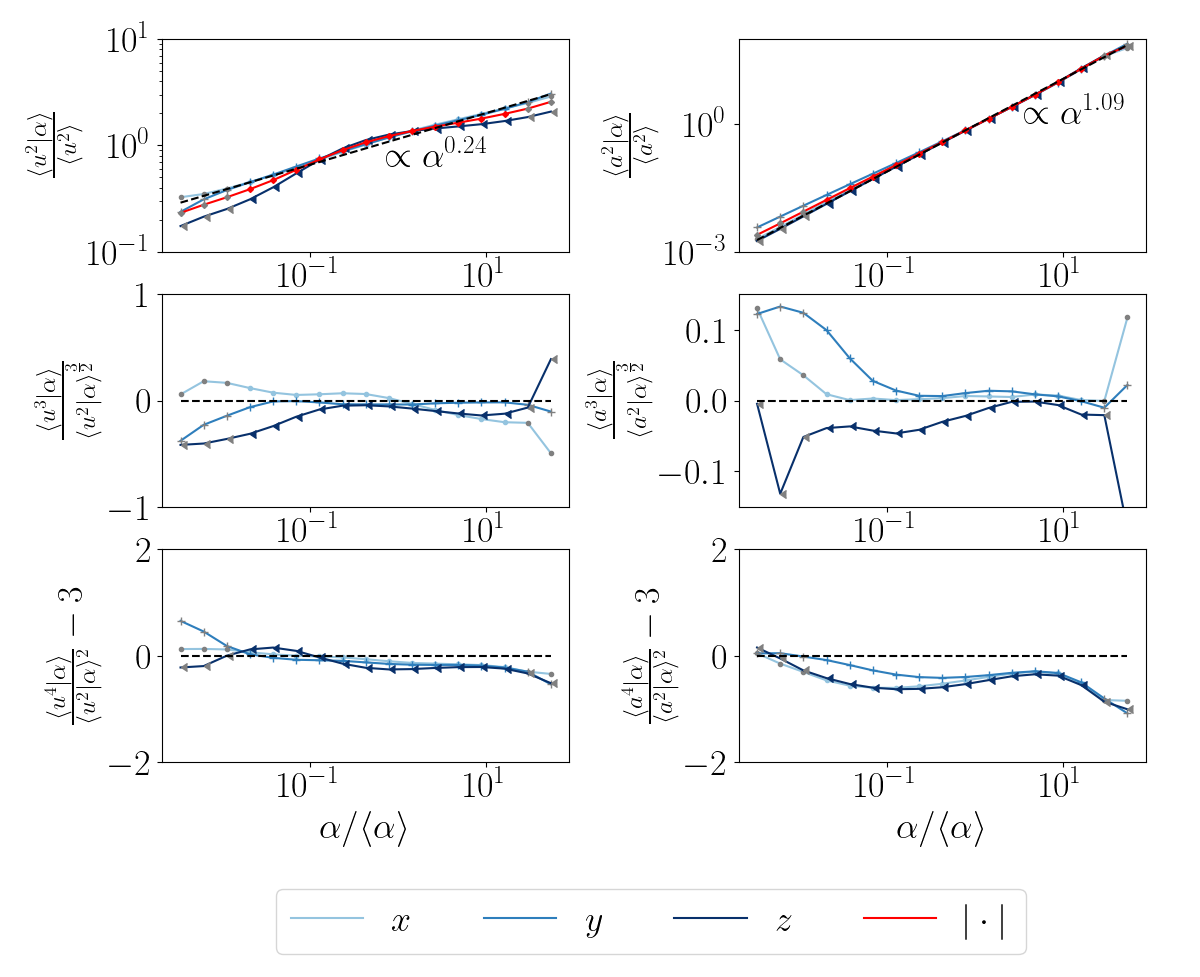}\\
    \vspace{-5em}
    Rayleigh-B\'enard convection \\
%von K\'arm\'an flow  
 \vspace{3em}
    \caption{Moments of velocity and acceleration in each subensembles as functions of the coarse-grained acceleration $\alpha$ for vK-flow (upper panels) and  
	    RBC (lower panels).
            Gray squares correspond to subensembles that have been generated but excluded from the analysis due to 
	    a lack of statistical convergence. The dashed lines in the respective top two panels show fits to the data.  
	    }
    \label{fig:moments}
\end{figure*}

\blue{To give quantitative information on the fluctuations of all velocity and acceleration components, fig.~\ref{fig:moments} shows the dependence of variance, skewness and
flatness of all velocity and acceleration components on $\alpha$ in each
subensemble for both datasets.  Subensembles marked in gray have been
excluded from the analysis due to a lack of statistical convergence and have not been included in any fits. As can be seen from the
data shown in the top panels, velocity and acceleration variances scale with
$\alpha$ for both von K\'arm\'an flow and Rayleigh-B\'enard convection
\begin{equation}
	\langle u_i^2 | \alpha \rangle / \langle u_i^2 \rangle  \propto \alpha^{\xi_u} \qquad \text{and} \qquad
	\langle a_i^2 | \alpha \rangle / \langle a_i^2 \rangle  \propto \alpha^{\xi_a} \ , 
\end{equation}
where $i \in {x,y,z}$ indicates the respective velocity and acceleration component, and similarly for    
\begin{equation}
	\langle |\bm{u}^2| | \alpha \rangle / \langle |\bm{u}^2| \rangle  \propto \alpha^{\xi_u} \qquad \text{and} \qquad
	\langle |\bm{a}^2| | \alpha \rangle / \langle |\bm{a}^2| \rangle  \propto \alpha^{\xi_a} \ . 
\end{equation}
For the acceleration, where very little deviation between the different
acceleration components is observed and the power-law of the acceleration
variance with $\alpha$ is very clear, we measure $\xi_a = 0.99 \pm 0.02$ for
von K\'arm\'an flow and $\xi_a = 1.09 \pm 0.04$ for Rayleigh-B\'enard
convection. For the velocity, where to a first approximation one would expect
$\xi_u = 0$ as the full velocity field itself has approximately Gaussian
statistics, we observe some deviations between the different spatial directions,
especially for von K\'arm\'an flow at low values of $\alpha$, which reflects
the presence of large-scale intermittency and anisotropy in this type of flow
\cite{Voth2002,Ouellette2006}.  The dependence of the velocity variance on
$\alpha$ is weak in this case, with $\xi_u = 0.14 \pm 0.03$. For Rayleigh-B\'enard
convection, we measure $\xi_u = 0.24 \pm 0.05$. 
The middle and bottom rows of panels presents skewness and flatness of velocity
and acceleration PDFs as function of $\alpha$ for both datasets. For the
acceleration, skewness values are generally small for data used in our
analyses, with $|S| < 0.05$ for all acceleration components. More variation
occurs in the velocity PDF skewness, with $|S| < 0.4$ for von K\'arm\'an flow and $|S| < 0.1$ for Rayleigh-B\'enard convection.
Flatness values are generally close to Gaussian, with $|\left < u^4 \right > /
\left < u^2 \right >^2 - 3| \leqslant 0.5 $   and $|\left < a^4 \right > /
\left < a^2 \right >^2 -3 |\leqslant 0.5 $, where $u$ and $a$ stand for any
velocity and acceleration component, respectively.} That is, skewness and flatness values for the velocity and acceleration component PDFs are close to Gaussian values %, with approximately zero skewness and flatness values of  $|\left < Y^4 \right > / \left < Y^2 \right >^2 - 3| \leqslant 0.5 $, where $Y$ stands for any velocity and acceleration component, respectively. This holds 
across both datasets and for all values of $\alpha$ considered.

% moments and scaling
%\begin{figure}[h!]

In summary, conditioning on the coarse-grained acceleration results in a decomposition 
of the full turbulent ensemble into approximately Gaussian subsensembles, as had been shown for 
DNS data of homogeneous and isotropic turbulence \cite{Bentkamp2019}, 
also for turbulent flows in the laboratory, here for vKF and RBC. % first main result -- applicability of the mechanism

\subsection{Curvature statistics}
\begin{figure}
	\centering
 \includegraphics[width = \columnwidth]{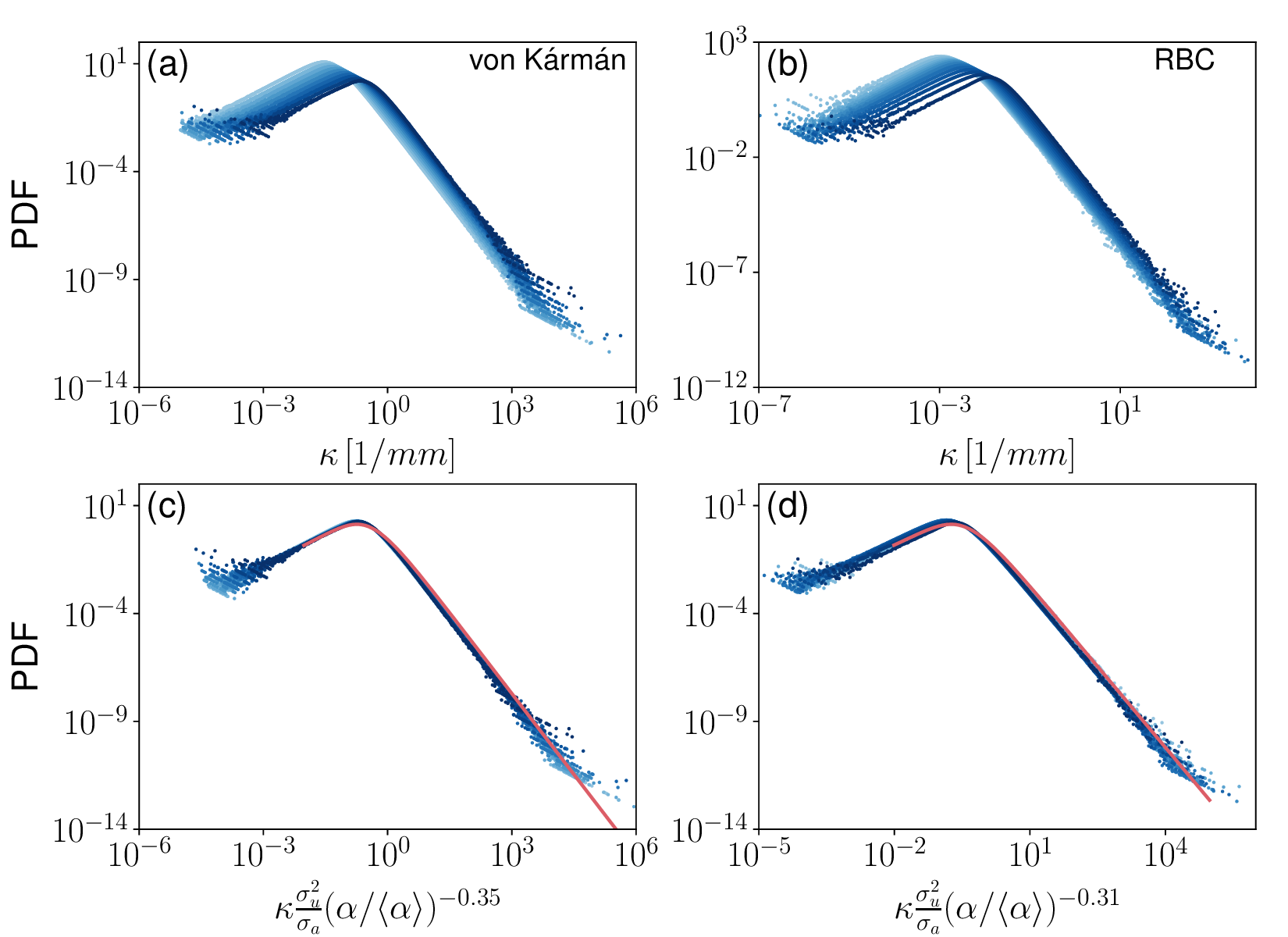}
    \caption{Curvature PDFs in Gaussian subensembles. 
	    (a) vKF, (b) RBC, (c) vKF, rescaled curvature PDFs, (d) RBC, rescaled curvature PDFs. 
	    The color gradient indicating increasing values of $\alpha$. The 
	    red line in panels (c) and (d) corresponds to Eq.~\eqref{eq:non-dim-pdf}.
	}
    \label{fig:pdf-curv-subsens}
\end{figure}

% subensembles, vK and RBC
Figures \ref{fig:pdf-curv-subsens} (a,b) present curvature PDFs
in each (approximately) Gaussian 
subensemble for vKF (a) and RBC (b). \blue{For both datasets, we observe that the curvature PDFs in each subensemble have a similar shape, which suggest that they can be collapsed onto} 
%The respective curves collapse on 
a master curve after re-scaling the curvature in each subensemble according to 
\begin{equation}
\kappa \propto \kappa \left(\sigma^{(\alpha)}_u \right)^2 / {\sigma^{(\alpha)}_a} 
	\propto \kappa \frac{\sigma_u^2}{\sigma_a} \frac{\alpha^{-\zeta}}{\langle \alpha \rangle^{-\zeta}} =  x_\alpha \ .  
    \label{eq:non-dim-curv}
\end{equation}
%where $\sigma^{(\alpha)}_u$ and $\sigma^{(\alpha)}_a$ are the standard deviations of velocity and acceleration, respectively, in a subensemble corresponding to a particular value of $\alpha$.
The exponent $\zeta \blue{= \xi_a/2 -\xi_u}$ is obtained by observing that $\sigma^{(\alpha)}_u$ and $\sigma^{(\alpha)}_a$ 
scale nontrivially with $\alpha$ \blue{as discussed in the previous section}. This implies that the variances of the near-Gaussian acceleration and 
velocity PDFs vary with $\alpha$, which encapsulates intermittency in terms of the absence of statistical self-similarity. For vK, we obtain $\zeta = 0.35 \pm 0.03$ and for RBC $\zeta = 0.31 \pm 0.05$.
% error propagation is additive as the exponents are subtracted and variances add up in this case, as u and a are uncorrelated in each subensemble. We state the standard error
Figures \ref{fig:pdf-curv-subsens} (c,d) present a comparison of the %obtained 
master curves 
with Eq.~\eqref{eq:non-dim-pdf} for vKF (c) and RBC (d). Model and 
data agree well, small deviations from the theoretical prediction are expected due to velocity and acceleration PDFs in each subensemble being not exactly Gaussian. 
%As can be seen from the data shown in 
%Fig.~\ref{fig:pdfs-subens}, the tails of especially the velocity component PDFs in each 
%subensemble are generally under-resolved. 
 
% final results
%
\section{Superstatistical descripton of curvature statistics }
% independence in subensembles 

\begin{figure*}
	\centering
    \vspace{-1em}
    \includegraphics[width = 0.9\columnwidth]{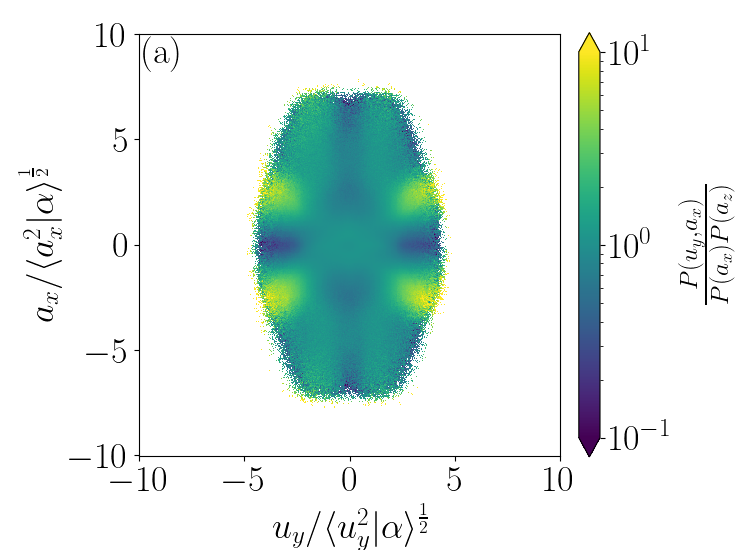} 
    \includegraphics[width = 0.9\columnwidth]{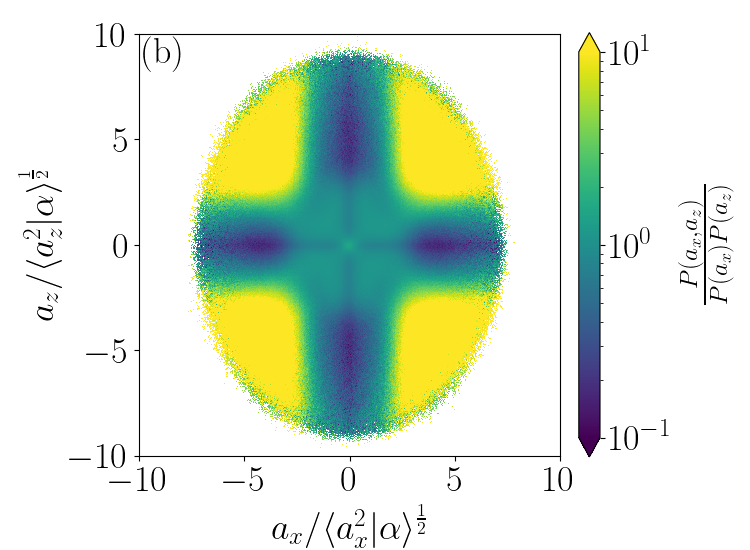} 
    \includegraphics[width = 0.9\columnwidth]{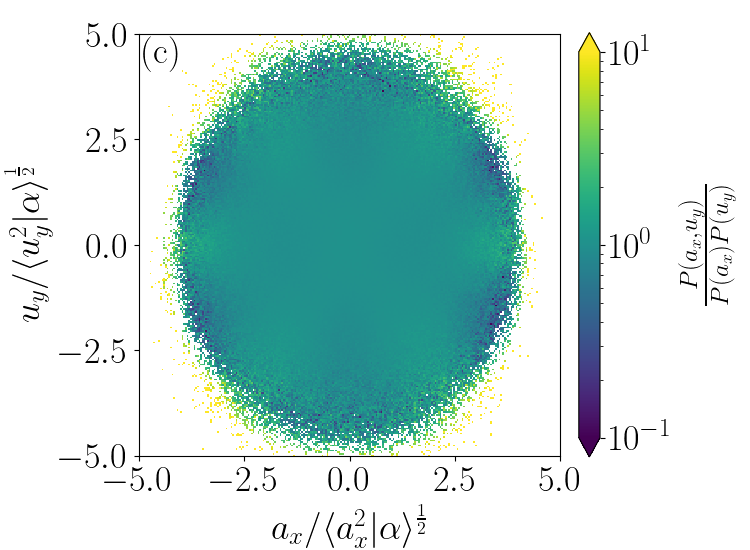}
    \includegraphics[width = 0.9\columnwidth]{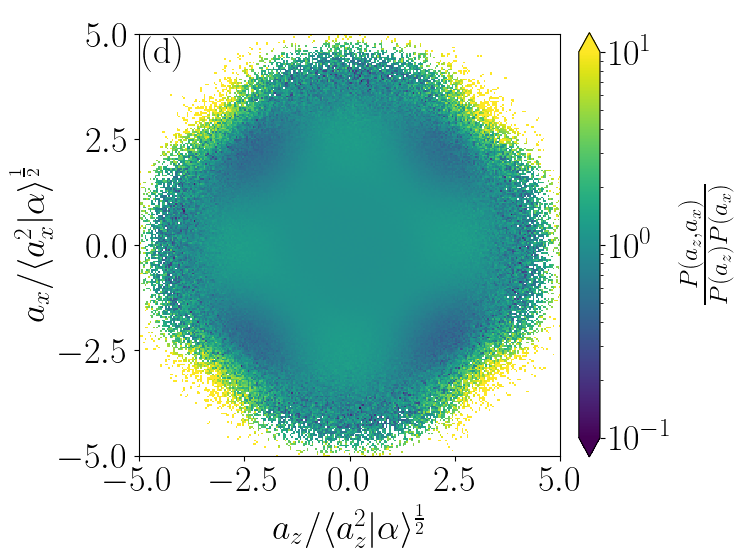}
    \caption{PDF quotients for vKF and RBC with $\alpha/\langle \alpha \rangle  \approx  2.6$. 
	    (a) vKF, velocity and acceleration components. 
	    (b) vKF, different acceleration components. 
	    (c) RBC, velocity and acceleration components. 
	    (d) RBC, different acceleration components. 
	    }
    \label{fig:jpdfs}
\end{figure*}

In order to apply the PDF model derived by Xu {\em et al.} \cite{Xu2007} \blue{in each approximately Gaussian subensemble}, it remains to be shown that the velocity and acceleration components in each subensemble are independent random variables. 
Uncorrelated and jointly Gaussian distributed random variables are statistically independent; therefore, we consider the quotient of the joint PDFs with the product of the marginal PDFs for velocity and acceleration components for both datasets.  
Figure 
\ref{fig:jpdfs} presents representative choices thereof.
As can be seen from the data, for vKF  
$u_y$ and $a_z$ are essentially uncorrelated, and
correlations between acceleration components $a_x$ and $a_z$ are significantly suppressed in the core of the joint PDF, compared with the full ensemble \cite{Mordant2004a}, however a level of correlation remains in the tails. For RBC, $u_y$ and $a_z$ are uncorrelated, and so are acceleration components $a_x$ and $a_z$. The remaining anisotropies and correlated regions in the joint PDFs mostly stem from the geometry of the large-scale forcing mechanisms, that is, the counter-rotating propellers in vKF and the LSC in RBC. 
Similar results hold for all components and all subensembles at least for the cores of the PDFs, a level of correlation usually remains in the tails for vK, which disappears under more conservative smoothing in the particle track reconstruction, as summarised in Appendix \ref{app:smoothing}. %As 
Furthermore, it can be shown analytically that velocity and acceleration components in the same direction % can be shown analytically to be 
are 
statistically independent, as a Gaussian random variable and 
its derivative are statistically uncorrelated \cite{Kallenberg}. 
%hence independent if they are jointly Gaussian distributed.
That is, the assumptions underlying the derivation of the model PDF are 
satisfied in each Gaussian subensemble, at least to a good approximation.

\begin{figure*}
	\centering
    \includegraphics[width = 1.8\columnwidth]{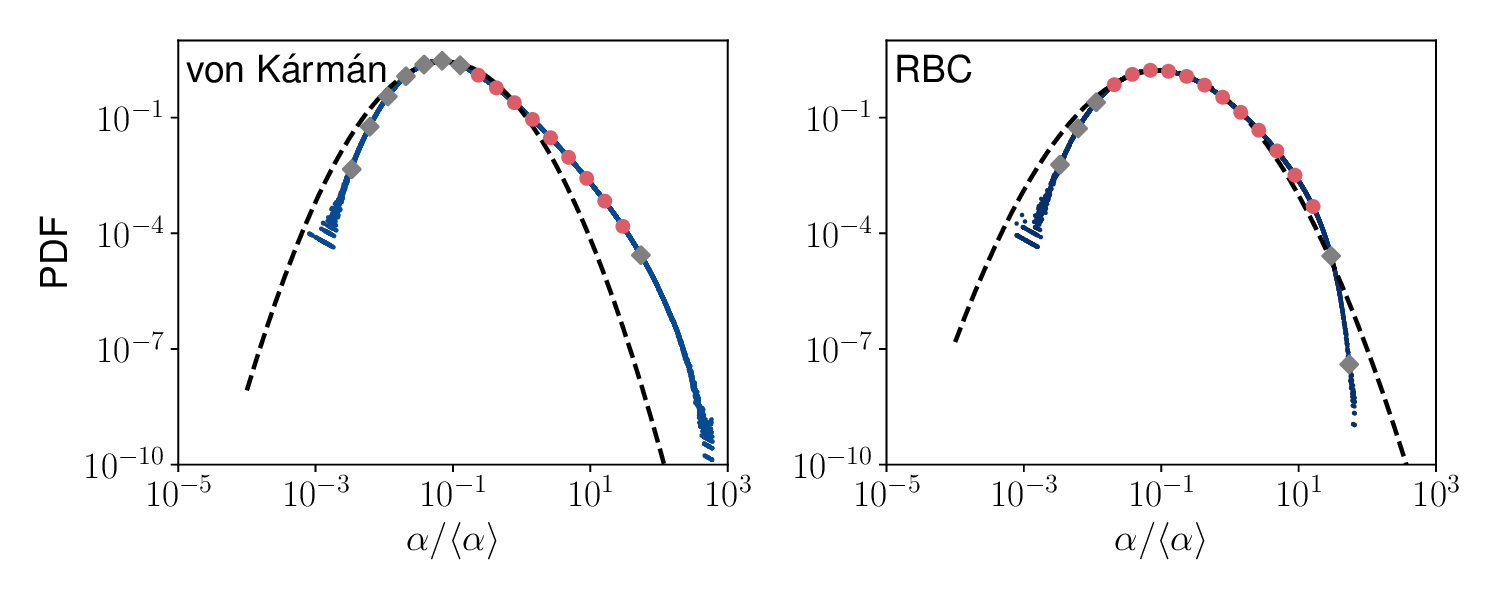}
    \caption{PDFs of the coarse-grained acceleration $\alpha$ for vK-flow (left) and RBC (right), calculated from trajectory data obtained with a crossover frequency set by assuming that the third derivative is noise. The red dots correspond to $\alpha$-values used to generate subensembles by conditioning. 
            Gray squares correspond to subensembles that have been generated but excluded from the analysis due to 
	    a lack of statistical convergence. The dashed lines show log-normal densities fitted to the core of the pdfs with parameters 
	      $(\mu, \sigma) = (-1.45 \pm 0.15, 1.06 \pm 0.11)$
        %$(\mu, \sigma) = (-1.42 \pm 0.04, 1.50 \pm 0.04)$ for 6 datapoints excluded, replace figure!
        for vK-flow  
        %in YH thesis (\mu, \sigma) = (-1.45 \pm 0.15, 1.06 \pm 0.11)$ using one less datapoint 
        and $(\mu, \sigma) = (-0.90 \pm 0.10,  1.20 \pm 0.07)$ for RBC. %[$(\mu, \sigma) = (-0.91 \pm 0.03, 1.33 \pm 0.03)$] 
	    }
    \label{fig:alpha-pdfs}
\end{figure*}

Having established that the assumptions justifying the derivation of the
closed-form expression for the curvature PDF \eqref{eq:non-dim-pdf} are valid to 
a good approximation 
in each subensemble for both datasets, the curvature of the full ensembles can
be constructed using the law of total probability \eqref{eq:reconstruction} by
rescaling the expression in \eqref{eq:non-dim-pdf} appropriately following
Eq.~\eqref{eq:non-dim-curv}. 
The required PDF of $\alpha$, $f(\alpha)$, is obtained either directly from
data or, to obtain a closed-form theoretical expression, 
by approximating $f(\alpha)$ with a log-normal distribution $f_{\rm model}(\alpha)$ with parameters $(\mu, \sigma)$  \cite{Bentkamp2019}. 
\blue{Figure
\ref{fig:alpha-pdfs} shows $f(\alpha)$ for vKF and RBC, compared with fitted log-normal distributions. As can be seen from the data, a log-normal density can only be fitted to data close to the mode,  both left and right tails are not captured. Good fits to the core of the respective PDFs are obtained for 
%$(\mu, \sigma) = (-1.42 \pm 0.04, 1.50 \pm 0.04)$ 
$(\mu, \sigma) = (-1.45 \pm 0.15, 1.06 \pm 0.11)$ %YH thesis
for vK, and $(\mu, \sigma) = (-0.90 \pm 0.10, 1.20 \pm 0.07)$ for RBC. % [$(\mu, \sigma) = (-0.91 \pm 0.03, 1.33 \pm 0.03)$ CHECK], 
%however, the tails
%are not captured well. %, especially for vKF 
For comparison, a log-normal density could be fitted to data obtained by direct numerical simulations of homogeneous and isotropic
turbulence \cite{Bentkamp2019} only for values larger than the mode of the $\alpha$-PDF including the right tail. 
We point out that the tails of the $\alpha$ PDF depend on the level of spatial and temporal resolution of the Lagrangian trajectories recorded in the experiment. As discussed in section \ref{sec:methods} B-splines of order three have been fitted to data, and the fitting procedure involves a smoothing parameter to account for the experimental uncertainty, the crossover frequency $f_c$.  Results obtained with a smaller crossover frequency $f_c = 0.06$ in all directions for von K\'arm\'an flow, that is, for a higher degree of smoothing, are presented in Appendix \ref{app:smoothing}, where the fit to a log-normal distribution is much better.      
}
As we shall see, a good approximation of the 
core %of $f(\alpha)$ 
is sufficient to obtain good agreement between theory and data.
 
\blue{
The symbols shown in both panels of fig.~\ref{fig:alpha-pdfs} correspond to $\alpha$-values used
to generate subensembles by conditioning. Red dots correspond to subsembles used in the analyses and gray squares correspond to
subensembles that have been generated but excluded from the analysis due to a
lack of statistical convergence. For von K\'arm\'an flow, ensembles near the mode have been excluded due to a very slow convergence of the velocity statistics in $y$-direction for small values of $\alpha$.} 

% \alpha pdfs
%\begin{figure}[h!]
%

% model
%\subsection{Closed-form approximation}
Before comparing theory to data, 
%using the scaling relation between the
%curvature PDFs in each Gaussian subensemble, 
we derive 
a closed-form model PDF
that includes the effects of intermittency as an extension of the Gaussian model \cite{Xu2007}.  
To do so, we \blue{consider the family of probability densities defined in eq.~\eqref{eq:model-pdf}, indexed by the values of the coarse-grained acceleration. Since the latter is a random conditioning observable, we can 
formally define
random variables $X:= X(\alpha)$ corresponding the non-dimensionalised curvature in each near-Gaussian} subensemble according to
Eq.~\eqref{eq:non-dim-curv}. Interpreting the set of possible $\alpha$-values as
an index set, which is possible as $\alpha\geq0$ \blue{and as the set of all $\alpha$-values is ordered, this} gives rise to a formal definition of a \blue{sequence of random variables }. We now relate the curvature
fluctuations to the expected behavior of this random \blue{sequence} %$X$ 
\begin{align}
\label{eq:expected_value}
    \langle X \rangle & = \int_0^{\infty} d\alpha f_{\rm model}(\alpha) X({\alpha}) \nonumber \\ 
	& = \int_0^{\infty} d\alpha f_{\rm model}(\alpha) \kappa \frac{\sigma_u^2}{\sigma_a} \frac{\alpha^{-\zeta}}{\langle \alpha \rangle^{-\zeta}} 
	= \kappa \frac{\sigma_u^2}{\sigma_a} \frac{\langle \alpha^{-\zeta} \rangle}{\langle \alpha \rangle^{-\zeta}} 
    \ ,
\end{align}
to approximate the PDF of $\kappa$ with that for the expected value of $X$
with respect to $A$, \blue{where we note that eq.~\eqref{eq:expected_value} provides a rescaling of the curvature with respect to an effective prefactor, $\langle \alpha^{-\zeta} \rangle/\langle \alpha \rangle^{-\zeta}$, resulting in}
%\blue{In this context, we interpret} the $x_{\alpha}$ \blue{to be} i.i.d. random variables, 
%\blue{Moreover,} the PDF of each $x_\alpha$ for a given value of $\alpha$ is known and given by 
%Eq.~\eqref{eq:non-dim-pdf}. %We now choose a particular value $\alpha^* \in A$, such that $\alpha^* = \langle \alpha^{-\zeta} \rangle$ 
%to obtain 
%ML: we don't need to choose alpha*, reformulate to say (5) provides an effective rescaling.  
%\yasmin{I think we went over it before but is this fine with units?} yes
\begin{equation} p_\kappa(\kappa) = \frac{\sigma_u^2}{\sigma_a}
	\frac{\langle \alpha^{-\zeta} \rangle}{\langle \alpha \rangle^{-\zeta}}
	p(\langle X \rangle) %= \frac{\sigma_u^2}{\sigma_a}
	%\frac{\langle \alpha^{-\zeta} \rangle}{\langle \alpha \rangle^{-\zeta}}
	%p(x_{\alpha^*}) 
    \ ,
\label{eq:model-pdf} 
\end{equation} 
where $p(\cdot)$ %$p(x_{\alpha^*})$ 
is given by Eq.~\eqref{eq:non-dim-pdf}.
In summary, we use the scaling of the acceleration and velocity variances to
map the full curvature PDF to that of a particular Gaussian subensemble, 
\blue{resulting in an effective closure. Finally, modelling the PDF of the coarse-grained acceleration as a log-normal distribution and using the expression for its moments in terms of its parameters $\mu$ and $\sigma$ results in the following analytical expression for the effective prefactor,
\begin{equation} 
\frac{\langle \alpha^{-\zeta} \rangle}{\langle \alpha \rangle^{-\zeta}}=\exp\!\left(\mu \zeta +\frac{1}{2}\sigma^2\zeta^2\right).
\end{equation} 
}

\begin{figure*}
	\centering 
    \vspace{-1.3em}
    \includegraphics[width = 0.9\columnwidth]{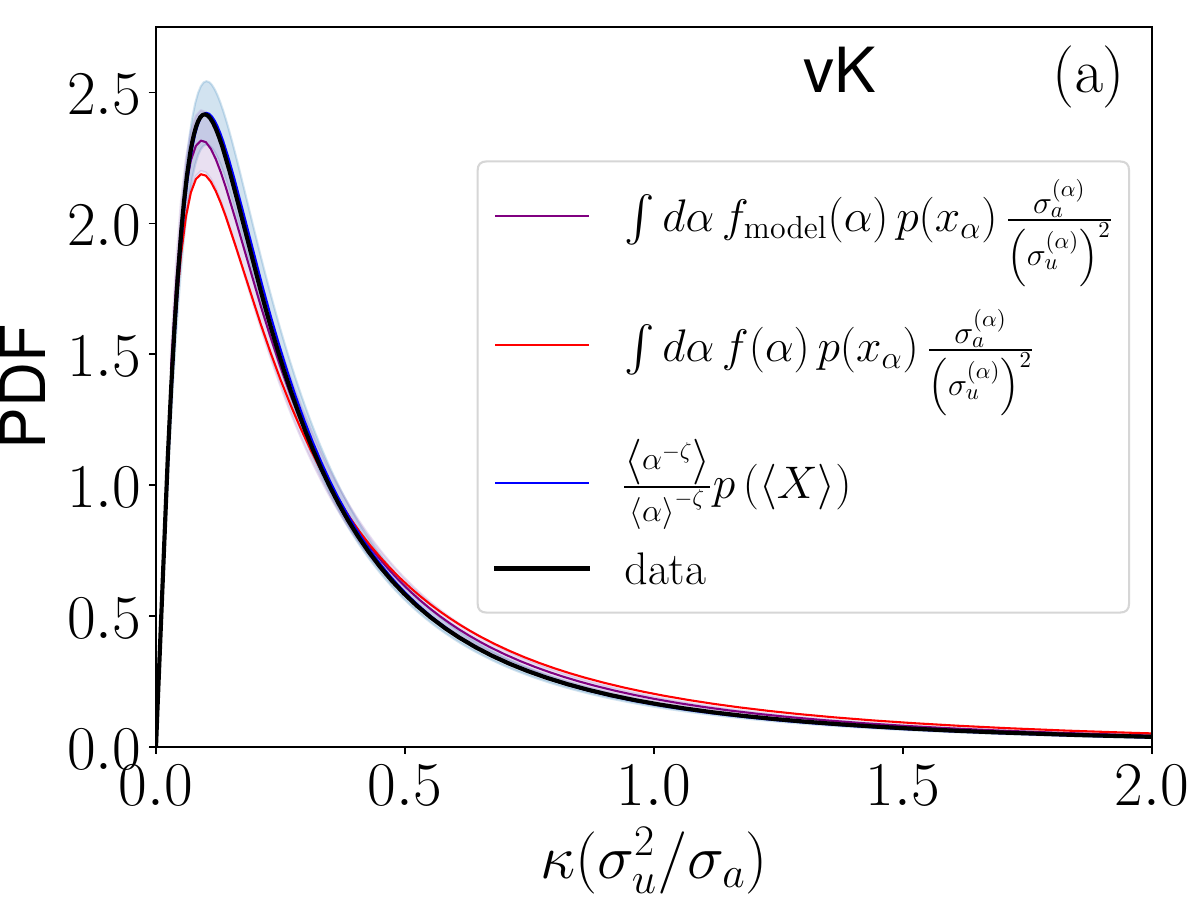}   \includegraphics[width = 0.9\columnwidth]{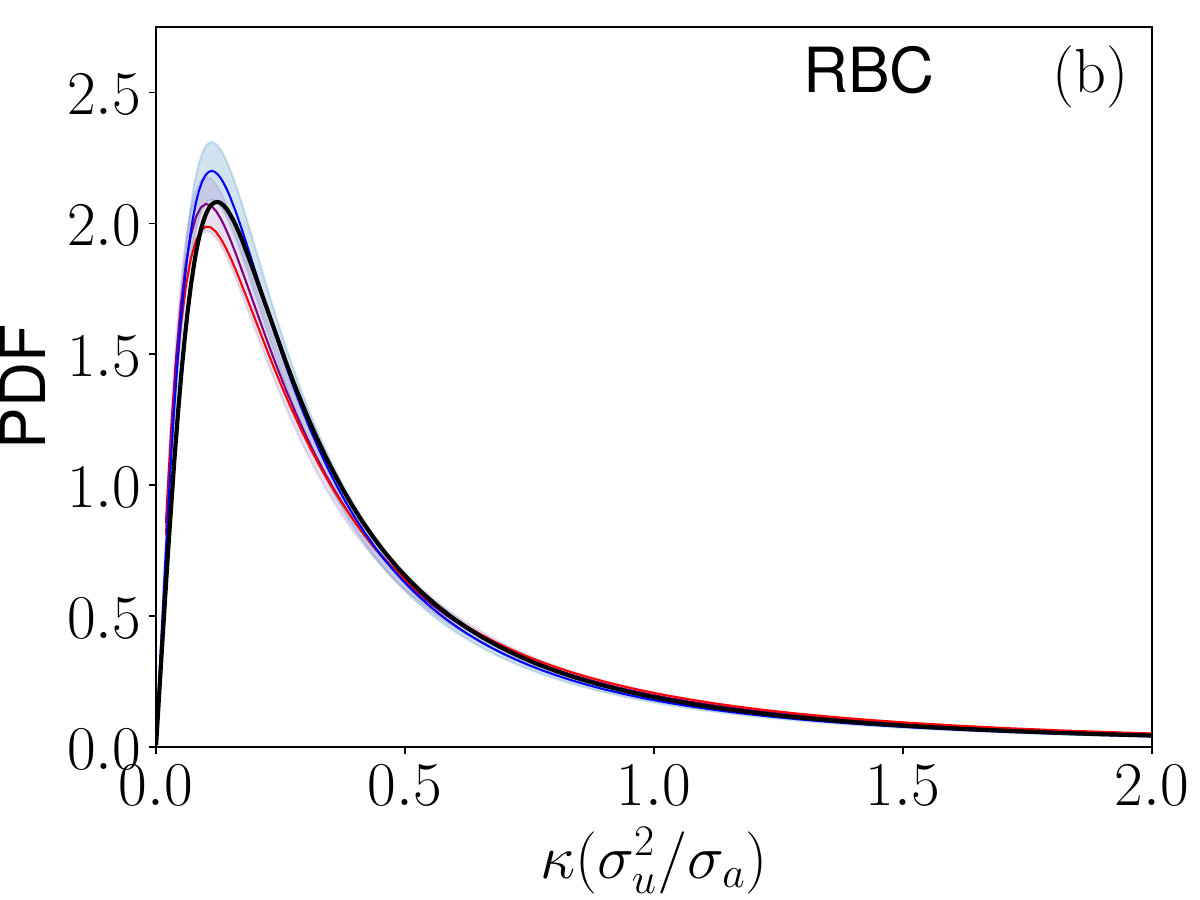}
    % same as YH-thesis fig 6.10
	\caption{Comparison between theory (Eq.~\eqref{eq:reconstruction}), with either the full $ \alpha-$PDF (red) or a fitted lognormal PDF (purple), model (Eq.~\eqref{eq:model-pdf}, blue) 
	and data (black) for (a) vKF, (b) RBC. %, with $c(\alpha) = \sigma^{(\alpha)}_a / \left(\sigma^{(\alpha)}_u \right)^2 $. 
    The shaded areas correspond to an error margin of $5 \%$.
	}
    \label{fig:pdf-curv}
\end{figure*}

%\subsection{Comparison to data}
%\label{sec:comparison}
Figure \ref{fig:pdf-curv} provides a comparison between the curvature PDF
calculated from the data shown in black, the theoretical prediction according
to Eq.~\eqref{eq:reconstruction} with either the full PDF of $\alpha$ or a fitted lognormal PDF shown in red and purple, respectively, 
and the model PDF \eqref{eq:model-pdf} shown in blue for vK
flow (a) and RBC (b). The shaded areas correspond to an uncertainty of $5 \%$.  As can be seen from the data presented in the
figure, theory and model capture generic behaviour and extreme events for both datasets. 
%agreement between theory, data and model is good for both datasets.
%The differences likely stem from the deviations between Gaussian PDFs and the data in each subensemble which, as disussed earlier, can be appreciated in Fig.~\ref{fig:pdfs-subens}.

% in fig, check if factors under integrals are correct, need to devide by sigma_u^2/sigma_a?

% 

% Discussion and conclusions
\section{Conclusions}
\label{sec:conclusions}
In summary, we present a closed-form model and an exact expression 
for the curvature PDF %for ensembles %\yasmin{should that be the full ensemble?}
of tracer particle trajectories in turbulence. 
%two turbulent flows, RBC and vKF, that includes intermittency effects. 
We combine a decomposition of the %respective 
full ensemble into near-Gaussian subensembles by statistical conditioning 
with an analytic form of the curvature PDF derived exactly for the 
case of Gaussian statistics, and reconstruct the PDF for the full ensemble by the law of total probability. Furthermore, a closed-form approximation for the curvature PDF is obtained by scaling the Gaussian
result with a correction factor that, as it involves moments of the coarse-grained acceleration, 
quantifies intermittency effects.  Both exact and approximate
expressions agree qualitatively and quantitatively with the curvature
PDFs sampled from tracer particle data from different turbulent flows. \blue{The present approach should be directly applicable to describe the PDF of the curvature angular momentum \cite{Braun2006} as well as that of the angle between subsequent particle displacement increments, at least in isotropic turbulence \cite{Bos2015}.}
%obtained from PIV measurements in von K\'arm\'an flow and Rayleigh-B\'enard convection.

Similar techniques should apply to a wider class 
of complex systems. For instance in plasma turbulence, 
%a comparison between data from numerical simulations and 
%the Xu {\em et al.} model \cite{Xu2007} suggest that 
a decomposition
into simpler subensembles may result in a full description of 
magnetic field-line statistics, at least at moderate levels
of anisotropy \cite{Luebke2024, Hengster2024Thesis}, \blue{and may provide further insight into the statistical properties of plasma impurities \cite{Kadoch_PoP_2022, Gheorghiu_2024, Lin_PPCF_2025}. }

%The same framework applies directly to curvature statistics in other intermittency-dominated systems, including plasma turbulence.

\begin{acknowledgments}
We thank M. Wilczek, L. Bentkamp, A. Morozov, and G. Vasil %, and K. Schneider 
%and F. Jacobitz % add later? 
for helpful discussions and J.
Weigel and M. Lellep for preliminary calculations.  Computational
resources on Cirrus ({\tt www.cirrus.ac.uk}) %\yasmin{I did most of the computations on archer for this project, shpuld we mention that/ replace cirrus?}
% not possible, as you piggy-backed on a different project on archer2
have been obtained through
Scottish Academic Access.  This work received funding from Priority
Programme SPP 1881 ``Turbulent Superstructures" of the Deutsche
Forschungsgemeinschaft (BO5544/1, LI3694/1,
SCHR1165/5), and the European High-Performance Infrastructures in Turbulence
(EuHIT) consortium for the DTrack measurement campaign at the von
K\'arm\'an flow facility GTF3. We acknowledge the support of the staff at MPIDS G{\"o}ttingen, in particular Eberhard Bodenschatz. 
%Yasmin Hengster was also supported by the School of Mathematics at the University of Edinburgh.
\end{acknowledgments}

%\clearpage

\appendix

\section{Smoothing}
\label{app:smoothing}
\blue{Here, we provide a comparison for vK-flow between data obtained for $f_c$ values of $(0.15,0.15,0.18)$ in $x$, $y$ and $z$-directions as in the main text and $f_c = 0.06$ in all directions, which results in smoother trajectories assuming higher experimental noise. }

\begin{figure}
	\centering
 \includegraphics[width = \columnwidth]{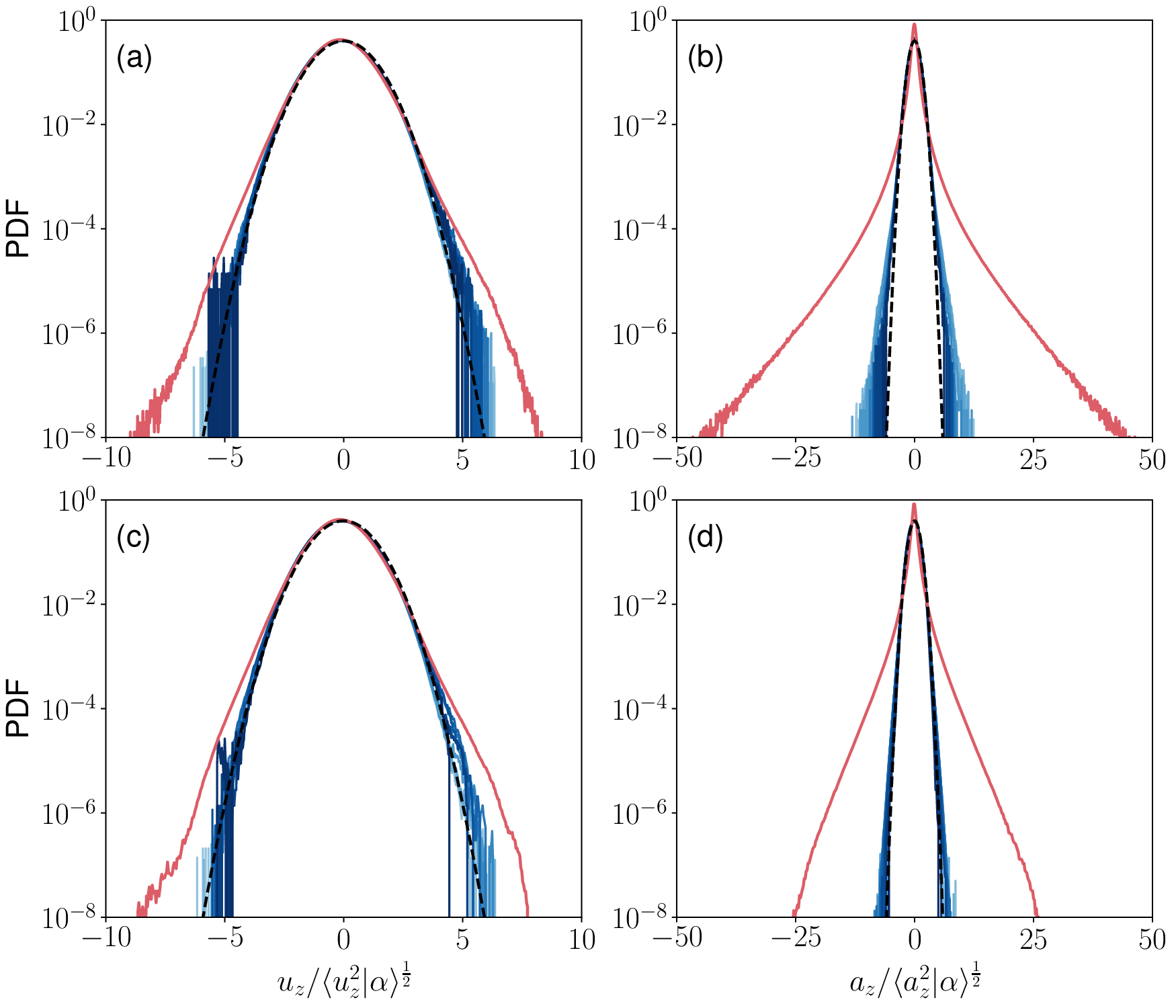}%{z-component_cf0.06.png}
 \caption{
 PDFs of the $z$-components of velocity (a,c) and acceleration (b,d) in vKF for the full ensemble (red) and in each subensemble (blue) for $f_c$ as in the main text (a,b) and $f_c = 0.06$ in all directions (c,d).  
	         The color gradient indicates increasing values of $\alpha$ as darker shades, the dashed line shows 
		 a Gaussian with zero mean and unity standard deviation.   
   }
 \label{fig:acceleration-pdfs-cf0.06}
\end{figure}

% vK velocity and acceleration pdfs
\blue{Figure \ref{fig:acceleration-pdfs-cf0.06} presents $z$-components velocity and acceleration PDF for the full ensemble and the subensembles obtained from fitting trajectories to data with the value of the crossover frequency used in the paper, $(0.15, 0.15, 0.18)$ in $x$, $y$ and $z$-directions, a lower crossover frequency $f_c = 0.06$ in all directions. As can be seen from a comparison of the velocity PDFs in subfigures (a) and (c), there is very little difference in the velocity statistics between the two datasets. In contrast, and as can be expected, the acceleration PDFs shown in subfigures (b) and (d) differ in the tails, especially for the full ensemble, with extreme acceleration events becoming much less likely with increased smoothing. For the subensembles, the Gaussian approximation of the PDFs holds better for the trajectories obtained with the $f_c = 0.06$ than for the data used in the manuscript, where the conditional PDFs retain slightly super-Gaussian tails. 
Acceleration PDFs of $x$- and $y$-components are similar to those shown in subfigures (b,d), and velocity PDFs $x$- and $y$-components almost Gaussian (not shown). The super-Gaussian tails in the $u_z$-PDF are related to the large-scale forcing in vKF.
}

%TODO -- remove replicated subfigs (a) and (b)

\begin{figure*}%[h!]
	\centering
    \includegraphics[width = 0.48\columnwidth]{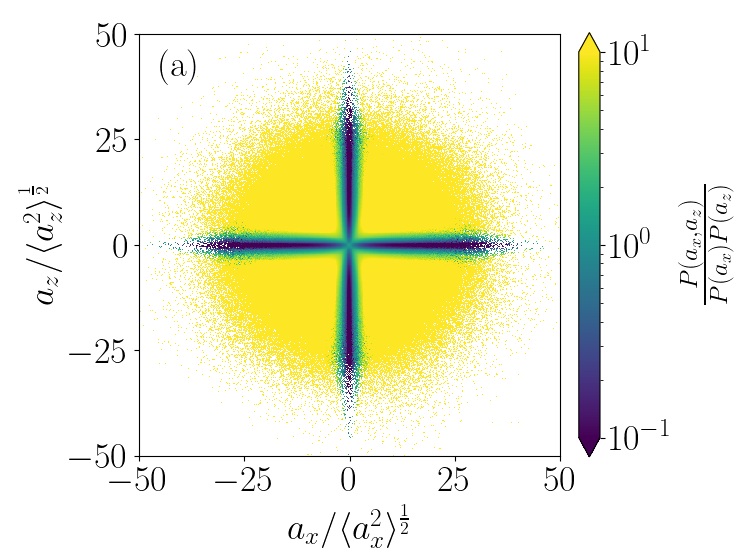}
    \includegraphics[width = 0.48\columnwidth]{vK_axaz_ens_11.png}
    \includegraphics[width = 0.48\columnwidth]{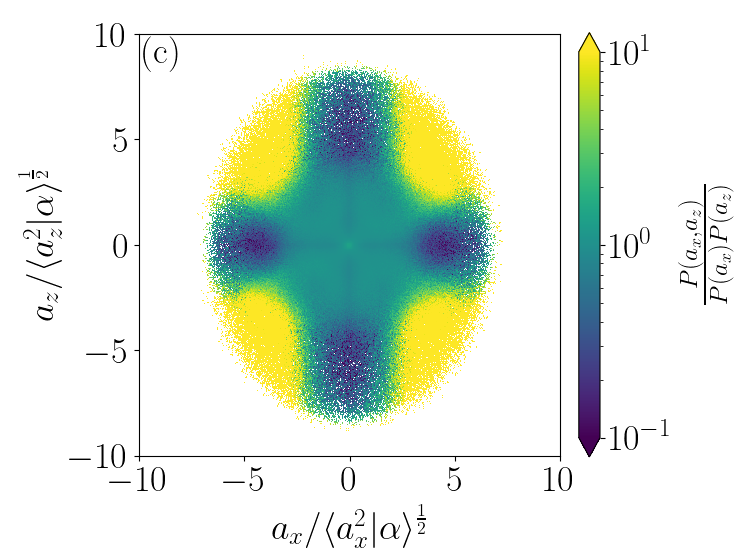}
    \includegraphics[width = 0.48\columnwidth]{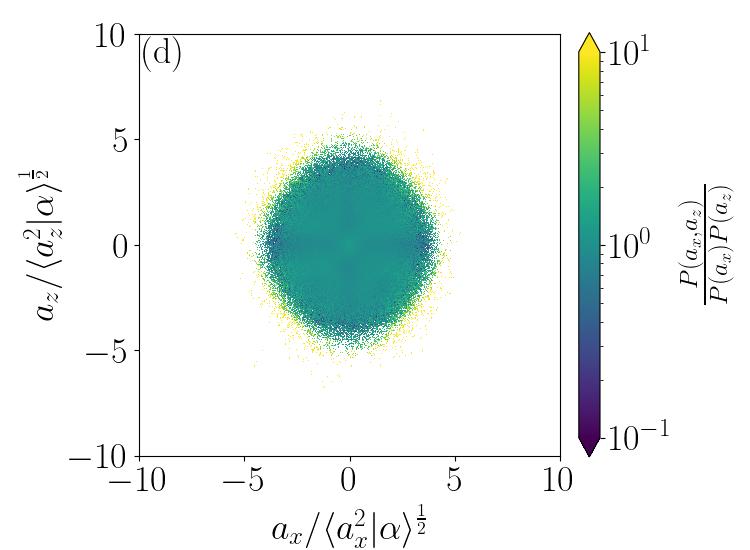}
    \includegraphics[width = 0.48\columnwidth]{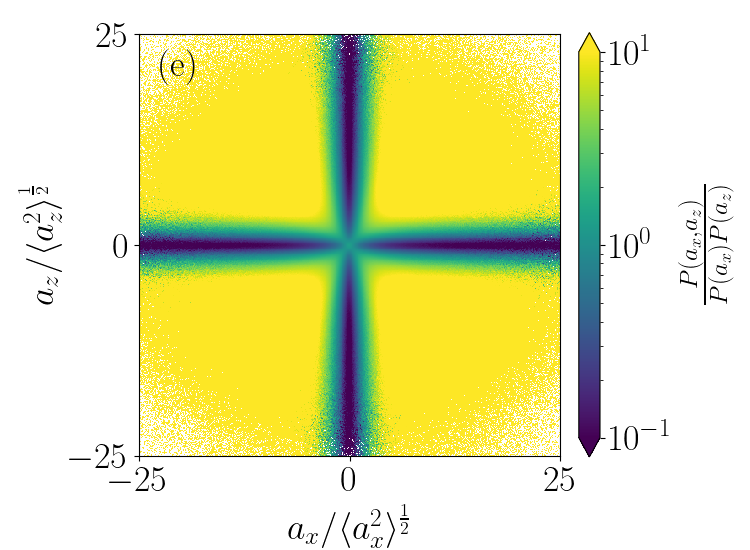}
    \includegraphics[width = 0.48\columnwidth]{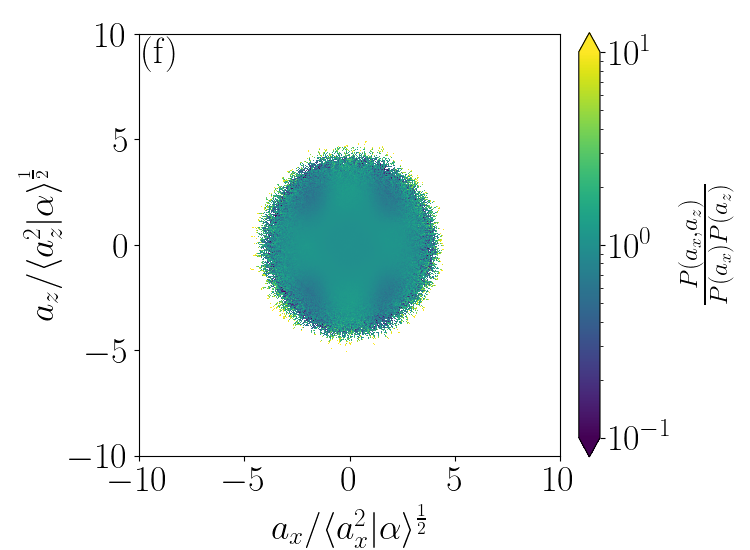}
    \includegraphics[width = 0.48\columnwidth]{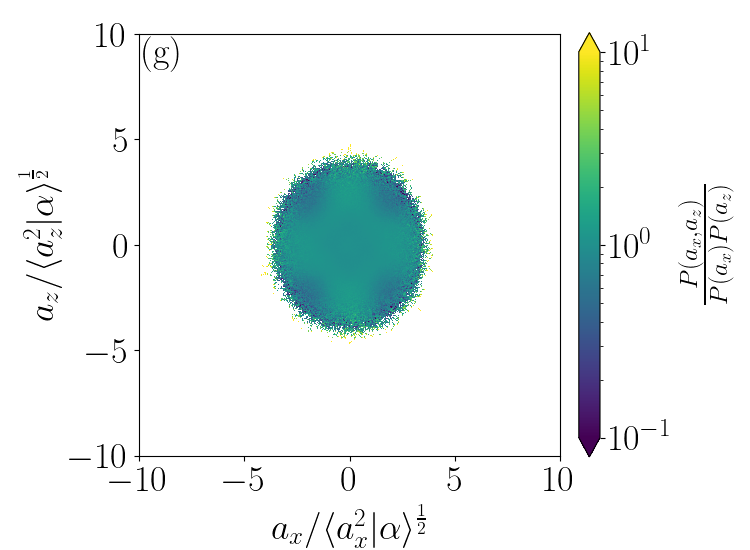}
    \includegraphics[width = 0.48\columnwidth]{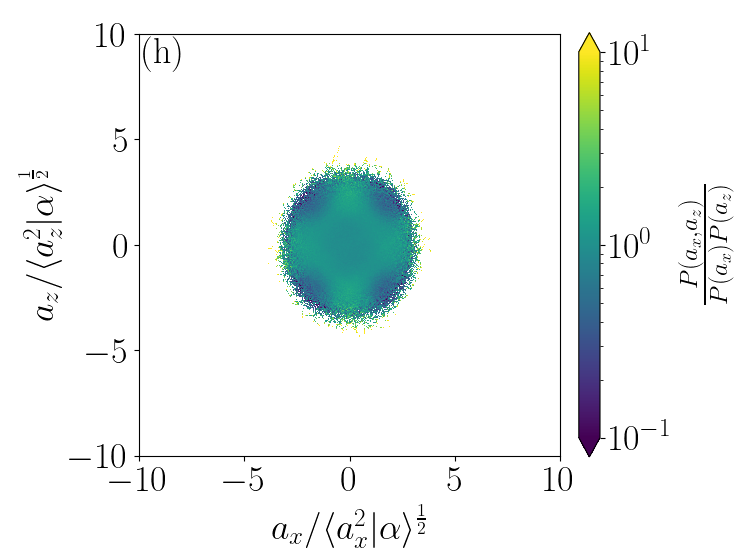}
    \caption{Acceleration PDF quotients for vK-flow for the full ensemble and different levels of $\alpha/\langle \alpha \rangle$ and different crossover frequencies. Top row (a-d) shows PDF quotients for a crossover frequency of $(0.15,0.15,0.18)$ in $x$, $y$ and $z$-direction, respectively as in the Letter, while the bottom row (e-g) shows results for a crossover frequency of $f_c = 0.06$ in all directions; 
	    (a,e) full ensemble, 
	    (b,f) $\alpha/\langle \alpha \rangle  \approx 2.6$,  
     (c,g) $\alpha/\langle \alpha \rangle  \approx  9$, 
     (d,h) $\alpha/\langle \alpha \rangle  \approx  30$. 
	    }
    \label{fig:jpdfs-app}
\end{figure*}

% vK pdf quotients 
\blue{Figure \ref{fig:jpdfs-app} presents the ratio between the joint PDF and the marginals for the full ensemble and several subensembles for vK-flow, calculated from trajectory data obtained with the value of the crossover frequency used in the paper, $(0.15,0.15,0.18)$ in $x$, $y$ and $z$-directions, (top row) and for a lower crossover frequency $f_c = 0.06$ in all directions, resulting in smoother trajectories. As can be seen from the data, correlations between acceleration components are significantly reduced in the subensembles compared with the full ensemble. However, for the value of the crossover frequency used in the analyses, for some ensembles correlations in the tails remain. These disappear if smoother trajectories are obtained by fitting with a higher value of the crossover frequency, as can be seen from the data shown in the bottom panels of fig.~\ref{fig:jpdfs-app}. By comparison of the PDFs in each subensembles between the data pertaining to the Letter (panels (b-d)) and that obtained with $f_c = 0.06$ (panels (f-h)), it can be seen that the smoothing suppresses the residual super-Gaussian tails of the conditional PDFs, where correlations remain. }

\begin{figure*}
	\centering
    \includegraphics[width = \columnwidth]{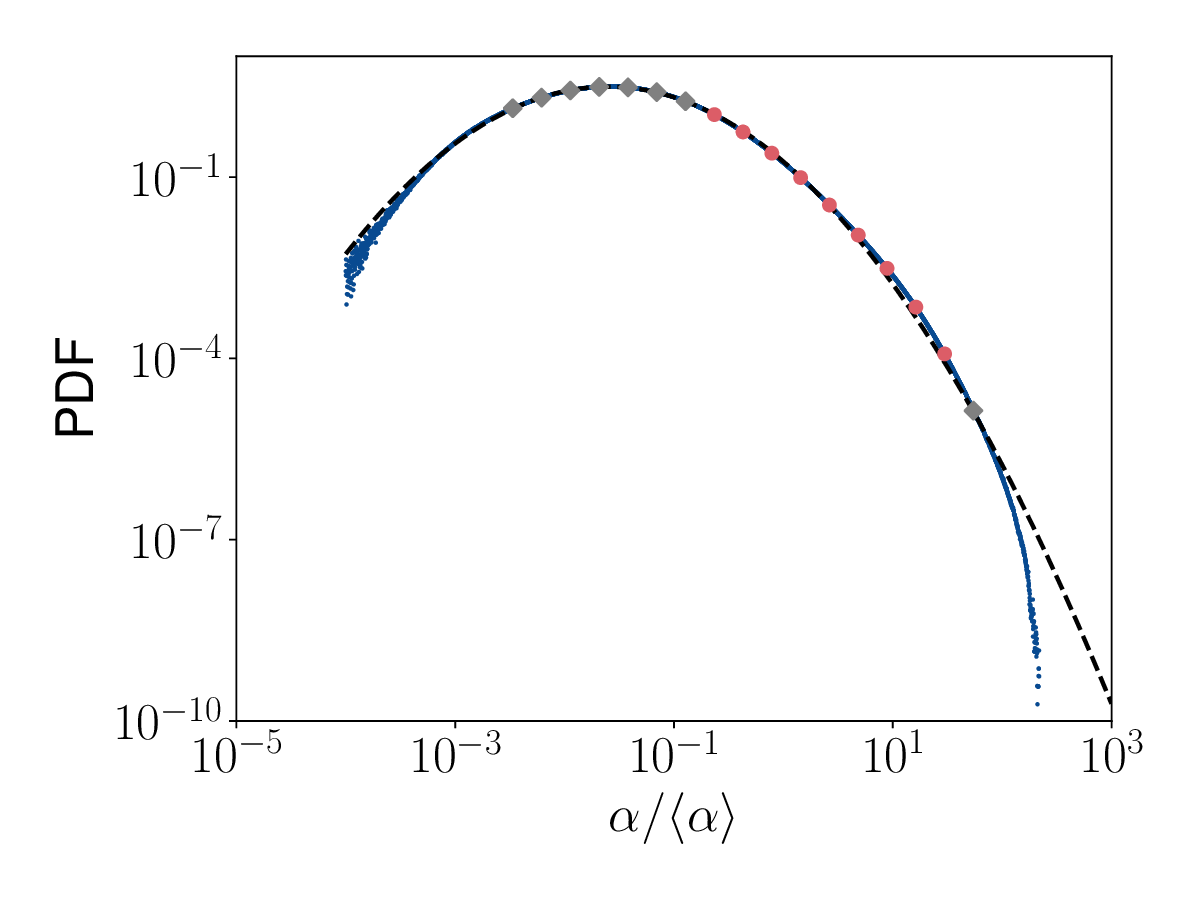}
    %{VK-alpha-pdf-fc0p06.png} %{alpha_PDF_cf_0.06.png}
    \includegraphics[width = 0.95\columnwidth]{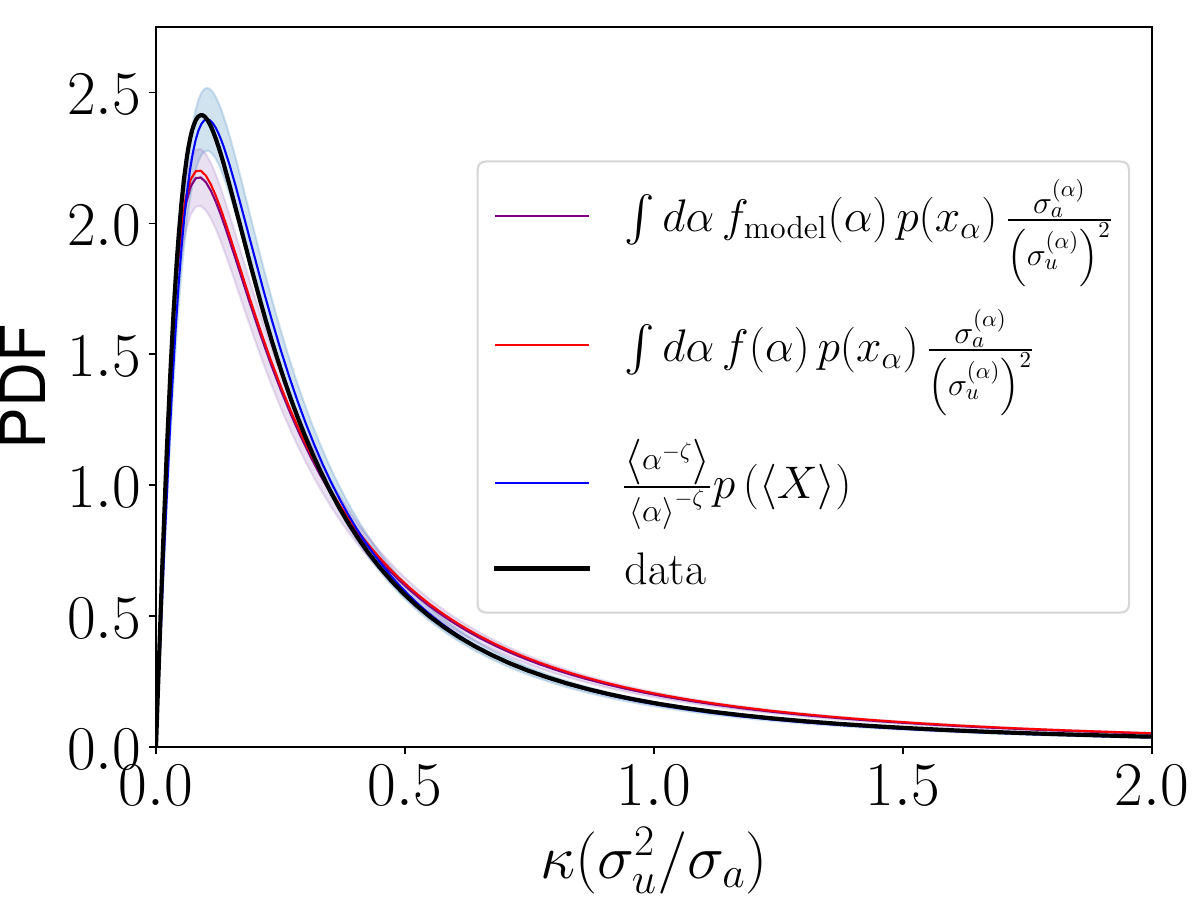}
    \caption{Left: PDFs of the coarse-grained acceleration $\alpha$ for vK-flow 
	       calculated from trajectory data obtained with a lower crossover frequency, $f_c = 0.06$ resulting in smoother fits. The red dots correspond to $\alpha$-values used to generate subensembles by conditioning. 
            Gray squares correspond to subensembles that have been generated but excluded from the analysis due to 
	    a lack of statistical convergence. The dashed lines show a log-normal density fitted to the core of the pdf with parameters 
	    $(\mu, \sigma) = (-1.30 \pm 0.03, 1.55 \pm 0.01)$.
        Right: Comparison between theory, with either the full PDF (red) or a fitted lognormal PDF (purple) has been used, model (blue) 
	and data (black) for vK-flow. Here, trajectories obtained with a crossover frequency of $f_c = 0.06$ have been used, corresponding to the data shown in the left panel and fig.~\ref{fig:jpdfs}, bottom row. The shaded areas correspond to an error margin of $5 \%$.
	    }
     \label{fig:alpha-pdfs-cf0.06}
\end{figure*}

% alpha-pdfs, final results
\blue{As can be expected, high levels of smoothing result in a increased likelihood of small values and a decreased likelihood of  high values of the coarse-grained acceleration, and {\em {vice versa}} for low levels of smoothing. As can be seen from the data presented in fig.~\ref{fig:alpha-pdfs-cf0.06} (a), for von K\'arm\'an flow with $f_c = 0.06$ the PDF can be approximated well with a log-normal
density except for extreme events in the far right tail whose likelihood is overestimated by the log-normal density. We point out that the final results, that is (i) the reconstruction of the curvature PDF of the full ensemble from those in each approximately Gaussian subensemble, and (ii) a model for the curvature PDF that includes intermittency effects, are insensitive to the changes in crossover frequency discussed here. This can be seen from the data shown in fig.~\ref{fig:alpha-pdfs-cf0.06}(b), where the curvature pdf sampled from data, its reconstructions using the sampled PDF of the coarse-grained acceleration and the fitted log-normal and the effective closed-form model are shown. Apart from a better agreement between the reconstructions, which is not surprising since the $f(\alpha)$ is much better approximated by a log-normal distribution for then higher level of smoothing considered here, the results shown in the right panel of fig.~\ref{fig:alpha-pdfs-cf0.06} are almost indistinguishable from those presented in the main text in fig.~\ref{fig:pdf-curv}}.

%$$ \beta^*=\exp\!\left(\mu-\frac{2}{3}\sigma^2\right). $$

\clearpage

\bibliographystyle{apsrev4-2}
\bibliography{apssamp}% Produces the bibliography via BibTeX.

%apsrev4-2.bst 2019-01-14 (MD) hand-edited version of apsrev4-1.bst
%Control: key (0)
%Control: author (72) initials jnrlst
%Control: editor formatted (1) identically to author
%Control: production of article title (-1) disabled
%Control: page (0) single
%Control: year (1) truncated
%Control: production of eprint (0) enabled
\providecommand{\noopsort}[1]{}\providecommand{\singleletter}[1]{#1}%
\begin{thebibliography}{47}%
\makeatletter
\providecommand \@ifxundefined [1]{%
 \@ifx{#1\undefined}
}%
\providecommand \@ifnum [1]{%
 \ifnum #1\expandafter \@firstoftwo
 \else \expandafter \@secondoftwo
 \fi
}%
\providecommand \@ifx [1]{%
 \ifx #1\expandafter \@firstoftwo
 \else \expandafter \@secondoftwo
 \fi
}%
\providecommand \natexlab [1]{#1}%
\providecommand \enquote  [1]{``#1''}%
\providecommand \bibnamefont  [1]{#1}%
\providecommand \bibfnamefont [1]{#1}%
\providecommand \citenamefont [1]{#1}%
\providecommand \href@noop [0]{\@secondoftwo}%
\providecommand \href [0]{\begingroup \@sanitize@url \@href}%
\providecommand \@href[1]{\@@startlink{#1}\@@href}%
\providecommand \@@href[1]{\endgroup#1\@@endlink}%
\providecommand \@sanitize@url [0]{\catcode `\\12\catcode `\$12\catcode
  `\&12\catcode `\#12\catcode `\^12\catcode `\_12\catcode `\%12\relax}%
\providecommand \@@startlink[1]{}%
\providecommand \@@endlink[0]{}%
\providecommand \url  [0]{\begingroup\@sanitize@url \@url }%
\providecommand \@url [1]{\endgroup\@href {#1}{\urlprefix }}%
\providecommand \urlprefix  [0]{URL }%
\providecommand \Eprint [0]{\href }%
\providecommand \doibase [0]{https://doi.org/}%
\providecommand \selectlanguage [0]{\@gobble}%
\providecommand \bibinfo  [0]{\@secondoftwo}%
\providecommand \bibfield  [0]{\@secondoftwo}%
\providecommand \translation [1]{[#1]}%
\providecommand \BibitemOpen [0]{}%
\providecommand \bibitemStop [0]{}%
\providecommand \bibitemNoStop [0]{.\EOS\space}%
\providecommand \EOS [0]{\spacefactor3000\relax}%
\providecommand \BibitemShut  [1]{\csname bibitem#1\endcsname}%
\let\auto@bib@innerbib\@empty
%</preamble>
\bibitem [{\citenamefont {Anderson}(2016)}]{Anderson2016}%
  \BibitemOpen
  \bibfield  {author} {\bibinfo {author} {\bibfnamefont {J.~D.}\ \bibnamefont
  {Anderson}},\ }\href@noop {} {\emph {\bibinfo {title} {Fundamentals of
  Aerodynamics}}},\ \bibinfo {edition} {6th}\ ed.\ (\bibinfo  {publisher}
  {McGraw--Hill Education},\ \bibinfo {address} {New York},\ \bibinfo {year}
  {2016})\BibitemShut {NoStop}%
\bibitem [{\citenamefont {Drazin}\ and\ \citenamefont
  {Reid}(2004)}]{DrazinReid2004}%
  \BibitemOpen
  \bibfield  {author} {\bibinfo {author} {\bibfnamefont {P.~G.}\ \bibnamefont
  {Drazin}}\ and\ \bibinfo {author} {\bibfnamefont {W.~H.}\ \bibnamefont
  {Reid}},\ }\href@noop {} {\emph {\bibinfo {title} {Hydrodynamic
  Stability}}},\ \bibinfo {edition} {2nd}\ ed.\ (\bibinfo  {publisher}
  {Cambridge University Press},\ \bibinfo {address} {Cambridge},\ \bibinfo
  {year} {2004})\BibitemShut {NoStop}%
\bibitem [{\citenamefont {Pakdel}\ and\ \citenamefont
  {McKinley}(1996)}]{PakdelMcKinley1996}%
  \BibitemOpen
  \bibfield  {author} {\bibinfo {author} {\bibfnamefont {P.}~\bibnamefont
  {Pakdel}}\ and\ \bibinfo {author} {\bibfnamefont {G.~H.}\ \bibnamefont
  {McKinley}},\ }\href {https://doi.org/10.1103/PhysRevLett.77.2459} {\bibfield
   {journal} {\bibinfo  {journal} {Phys.\ Rev.\ Lett.}\ }\textbf {\bibinfo
  {volume} {77}},\ \bibinfo {pages} {2459} (\bibinfo {year}
  {1996})}\BibitemShut {NoStop}%
\bibitem [{\citenamefont {Moffatt}(1978)}]{Moffatt1978}%
  \BibitemOpen
  \bibfield  {author} {\bibinfo {author} {\bibfnamefont {H.~K.}\ \bibnamefont
  {Moffatt}},\ }\href@noop {} {\emph {\bibinfo {title} {Magnetic Field
  Generation in Electrically Conducting Fluids}}}\ (\bibinfo  {publisher}
  {Cambridge University Press},\ \bibinfo {address} {Cambridge},\ \bibinfo
  {year} {1978})\BibitemShut {NoStop}%
\bibitem [{\citenamefont {Drake}\ \emph {et~al.}(2006)\citenamefont {Drake},
  \citenamefont {Swisdak}, \citenamefont {Che},\ and\ \citenamefont
  {Shay}}]{Drake2006}%
  \BibitemOpen
  \bibfield  {author} {\bibinfo {author} {\bibfnamefont {J.~F.}\ \bibnamefont
  {Drake}}, \bibinfo {author} {\bibfnamefont {M.}~\bibnamefont {Swisdak}},
  \bibinfo {author} {\bibfnamefont {H.}~\bibnamefont {Che}},\ and\ \bibinfo
  {author} {\bibfnamefont {M.~A.}\ \bibnamefont {Shay}},\ }\href
  {https://doi.org/10.1038/nature05116} {\bibfield  {journal} {\bibinfo
  {journal} {Nature}\ }\textbf {\bibinfo {volume} {443}},\ \bibinfo {pages}
  {553} (\bibinfo {year} {2006})}\BibitemShut {NoStop}%
\bibitem [{\citenamefont {Dahlin}\ \emph {et~al.}(2014)\citenamefont {Dahlin},
  \citenamefont {Drake},\ and\ \citenamefont {Swisdak}}]{Dahlin2014}%
  \BibitemOpen
  \bibfield  {author} {\bibinfo {author} {\bibfnamefont {J.~T.}\ \bibnamefont
  {Dahlin}}, \bibinfo {author} {\bibfnamefont {J.~F.}\ \bibnamefont {Drake}},\
  and\ \bibinfo {author} {\bibfnamefont {M.}~\bibnamefont {Swisdak}},\ }\href
  {https://doi.org/10.1063/1.4894484} {\bibfield  {journal} {\bibinfo
  {journal} {Phys. Plasmas}\ }\textbf {\bibinfo {volume} {21}},\ \bibinfo
  {pages} {092304} (\bibinfo {year} {2014})}\BibitemShut {NoStop}%
\bibitem [{\citenamefont {Braun}\ \emph {et~al.}(2006)\citenamefont {Braun},
  \citenamefont {Lillo},\ and\ \citenamefont {Eckhardt}}]{Braun2006}%
  \BibitemOpen
  \bibfield  {author} {\bibinfo {author} {\bibfnamefont {W.}~\bibnamefont
  {Braun}}, \bibinfo {author} {\bibfnamefont {F.~D.}\ \bibnamefont {Lillo}},\
  and\ \bibinfo {author} {\bibfnamefont {B.}~\bibnamefont {Eckhardt}},\
  }\href@noop {} {\bibfield  {journal} {\bibinfo  {journal} {J. Turbul.}\
  }\textbf {\bibinfo {volume} {7}},\ \bibinfo {pages} {N62} (\bibinfo {year}
  {2006})}\BibitemShut {NoStop}%
\bibitem [{\citenamefont {Scagliarini}(2011)}]{Scagliarini2011}%
  \BibitemOpen
  \bibfield  {author} {\bibinfo {author} {\bibfnamefont {A.}~\bibnamefont
  {Scagliarini}},\ }\href@noop {} {\bibfield  {journal} {\bibinfo  {journal}
  {J. Turbul.}\ }\textbf {\bibinfo {volume} {12}},\ \bibinfo {pages} {N25}
  (\bibinfo {year} {2011})}\BibitemShut {NoStop}%
\bibitem [{\citenamefont {Xu}\ \emph {et~al.}(2007)\citenamefont {Xu},
  \citenamefont {Ouellette},\ and\ \citenamefont {Bodenschatz}}]{Xu2007}%
  \BibitemOpen
  \bibfield  {author} {\bibinfo {author} {\bibfnamefont {H.}~\bibnamefont
  {Xu}}, \bibinfo {author} {\bibfnamefont {N.}~\bibnamefont {Ouellette}},\ and\
  \bibinfo {author} {\bibfnamefont {E.}~\bibnamefont {Bodenschatz}},\
  }\href@noop {} {\bibfield  {journal} {\bibinfo  {journal} {Phys.\ Rev.\
  Lett.}\ }\textbf {\bibinfo {volume} {98}},\ \bibinfo {pages} {050201}
  (\bibinfo {year} {2007})}\BibitemShut {NoStop}%
\bibitem [{\citenamefont {Hengster}\ \emph {et~al.}(2024)\citenamefont
  {Hengster}, \citenamefont {Lellep}, \citenamefont {Weigel}, \citenamefont
  {Bross}, \citenamefont {Bosbach}, \citenamefont {Schanz}, \citenamefont
  {Schr\"oder}, \citenamefont {Huhn}, \citenamefont {Novara}, \citenamefont
  {Garaboa~Paz}, \citenamefont {K\"ahler},\ and\ \citenamefont
  {Linkmann}}]{Hengster2023}%
  \BibitemOpen
  \bibfield  {author} {\bibinfo {author} {\bibfnamefont {Y.}~\bibnamefont
  {Hengster}}, \bibinfo {author} {\bibfnamefont {M.}~\bibnamefont {Lellep}},
  \bibinfo {author} {\bibfnamefont {J.}~\bibnamefont {Weigel}}, \bibinfo
  {author} {\bibfnamefont {M.}~\bibnamefont {Bross}}, \bibinfo {author}
  {\bibfnamefont {J.}~\bibnamefont {Bosbach}}, \bibinfo {author} {\bibfnamefont
  {D.}~\bibnamefont {Schanz}}, \bibinfo {author} {\bibfnamefont
  {A.}~\bibnamefont {Schr\"oder}}, \bibinfo {author} {\bibfnamefont
  {F.}~\bibnamefont {Huhn}}, \bibinfo {author} {\bibfnamefont {M.}~\bibnamefont
  {Novara}}, \bibinfo {author} {\bibfnamefont {D.}~\bibnamefont {Garaboa~Paz}},
  \bibinfo {author} {\bibfnamefont {C.}~\bibnamefont {K\"ahler}},\ and\
  \bibinfo {author} {\bibfnamefont {M.}~\bibnamefont {Linkmann}},\ }\href@noop
  {} {\bibfield  {journal} {\bibinfo  {journal} {Eur. J. Mech. B/Fluids}\
  }\textbf {\bibinfo {volume} {103}},\ \bibinfo {pages} {284} (\bibinfo {year}
  {2024})}\BibitemShut {NoStop}%
\bibitem [{\citenamefont {Alards}\ \emph {et~al.}(2017)\citenamefont {Alards},
  \citenamefont {Rajaei}, \citenamefont {Del~Castello}, \citenamefont {Kunnen},
  \citenamefont {Toschi},\ and\ \citenamefont {Clercx}}]{Alards2017}%
  \BibitemOpen
  \bibfield  {author} {\bibinfo {author} {\bibfnamefont {K.~M.~J.}\
  \bibnamefont {Alards}}, \bibinfo {author} {\bibfnamefont {H.}~\bibnamefont
  {Rajaei}}, \bibinfo {author} {\bibfnamefont {L.}~\bibnamefont
  {Del~Castello}}, \bibinfo {author} {\bibfnamefont {R.~P.~J.}\ \bibnamefont
  {Kunnen}}, \bibinfo {author} {\bibfnamefont {F.}~\bibnamefont {Toschi}},\
  and\ \bibinfo {author} {\bibfnamefont {H.~J.~H.}\ \bibnamefont {Clercx}},\
  }\href@noop {} {\bibfield  {journal} {\bibinfo  {journal} {Phys.\ Rev.\
  Fluids}\ }\textbf {\bibinfo {volume} {2}},\ \bibinfo {pages} {044601}
  (\bibinfo {year} {2017})}\BibitemShut {NoStop}%
\bibitem [{\citenamefont {Schekochihin}\ \emph {et~al.}(2001)\citenamefont
  {Schekochihin}, \citenamefont {Cowley}, \citenamefont {Maron},\ and\
  \citenamefont {Malyshkin}}]{Schekochihin2001}%
  \BibitemOpen
  \bibfield  {author} {\bibinfo {author} {\bibfnamefont {A.}~\bibnamefont
  {Schekochihin}}, \bibinfo {author} {\bibfnamefont {S.}~\bibnamefont
  {Cowley}}, \bibinfo {author} {\bibfnamefont {J.}~\bibnamefont {Maron}},\ and\
  \bibinfo {author} {\bibfnamefont {L.}~\bibnamefont {Malyshkin}},\ }\href@noop
  {} {\bibfield  {journal} {\bibinfo  {journal} {Phys. Rev. E}\ }\textbf
  {\bibinfo {volume} {65}},\ \bibinfo {pages} {016305} (\bibinfo {year}
  {2001})}\BibitemShut {NoStop}%
\bibitem [{\citenamefont {Schekochihin}\ \emph {et~al.}(2002)\citenamefont
  {Schekochihin}, \citenamefont {Maron}, \citenamefont {Cowley},\ and\
  \citenamefont {McWilliams}}]{Schekochihin2002}%
  \BibitemOpen
  \bibfield  {author} {\bibinfo {author} {\bibfnamefont {A.~A.}\ \bibnamefont
  {Schekochihin}}, \bibinfo {author} {\bibfnamefont {J.~L.}\ \bibnamefont
  {Maron}}, \bibinfo {author} {\bibfnamefont {S.~C.}\ \bibnamefont {Cowley}},\
  and\ \bibinfo {author} {\bibfnamefont {J.~C.}\ \bibnamefont {McWilliams}},\
  }\href@noop {} {\bibfield  {journal} {\bibinfo  {journal} {Astrophys.\ J.}\
  }\textbf {\bibinfo {volume} {576}},\ \bibinfo {pages} {806} (\bibinfo {year}
  {2002})}\BibitemShut {NoStop}%
\bibitem [{\citenamefont {Schekochihin}\ \emph {et~al.}(2004)\citenamefont
  {Schekochihin}, \citenamefont {Cowley}, \citenamefont {Taylor}, \citenamefont
  {Maron},\ and\ \citenamefont {McWilliams}}]{Schekochihin2004}%
  \BibitemOpen
  \bibfield  {author} {\bibinfo {author} {\bibfnamefont {A.~A.}\ \bibnamefont
  {Schekochihin}}, \bibinfo {author} {\bibfnamefont {S.~C.}\ \bibnamefont
  {Cowley}}, \bibinfo {author} {\bibfnamefont {S.~F.}\ \bibnamefont {Taylor}},
  \bibinfo {author} {\bibfnamefont {J.~L.}\ \bibnamefont {Maron}},\ and\
  \bibinfo {author} {\bibfnamefont {J.~C.}\ \bibnamefont {McWilliams}},\
  }\href@noop {} {\bibfield  {journal} {\bibinfo  {journal} {Astrophys.\ J.}\
  }\textbf {\bibinfo {volume} {612}},\ \bibinfo {pages} {276} (\bibinfo {year}
  {2004})}\BibitemShut {NoStop}%
\bibitem [{\citenamefont {Yang}\ \emph {et~al.}(2019)\citenamefont {Yang},
  \citenamefont {Wan}, \citenamefont {Matthaeus}, \citenamefont {Shi},
  \citenamefont {Parashar}, \citenamefont {Lu},\ and\ \citenamefont
  {Chen}}]{Yang2019}%
  \BibitemOpen
  \bibfield  {author} {\bibinfo {author} {\bibfnamefont {Y.}~\bibnamefont
  {Yang}}, \bibinfo {author} {\bibfnamefont {M.}~\bibnamefont {Wan}}, \bibinfo
  {author} {\bibfnamefont {W.~H.}\ \bibnamefont {Matthaeus}}, \bibinfo {author}
  {\bibfnamefont {Y.}~\bibnamefont {Shi}}, \bibinfo {author} {\bibfnamefont
  {T.~N.}\ \bibnamefont {Parashar}}, \bibinfo {author} {\bibfnamefont
  {Q.}~\bibnamefont {Lu}},\ and\ \bibinfo {author} {\bibfnamefont
  {S.}~\bibnamefont {Chen}},\ }\href@noop {} {\bibfield  {journal} {\bibinfo
  {journal} {Phys.~Plasmas}\ }\textbf {\bibinfo {volume} {26}},\ \bibinfo
  {pages} {072306} (\bibinfo {year} {2019})}\BibitemShut {NoStop}%
\bibitem [{\citenamefont {Bandyopadhyay}\ \emph {et~al.}(2020)\citenamefont
  {Bandyopadhyay}, \citenamefont {Yang}, \citenamefont {Matthaeus},
  \citenamefont {Chasapis}, \citenamefont {Parashar}, \citenamefont {Russell},
  \citenamefont {Strangeway}, \citenamefont {Torbert}, \citenamefont {Giles},
  \citenamefont {Gershman}, \citenamefont {Pollock}, \citenamefont {Moore},\
  and\ \citenamefont {Burch}}]{Bandyopadhyay2020}%
  \BibitemOpen
  \bibfield  {author} {\bibinfo {author} {\bibfnamefont {R.}~\bibnamefont
  {Bandyopadhyay}}, \bibinfo {author} {\bibfnamefont {Y.}~\bibnamefont {Yang}},
  \bibinfo {author} {\bibfnamefont {W.~H.}\ \bibnamefont {Matthaeus}}, \bibinfo
  {author} {\bibfnamefont {A.}~\bibnamefont {Chasapis}}, \bibinfo {author}
  {\bibfnamefont {T.~N.}\ \bibnamefont {Parashar}}, \bibinfo {author}
  {\bibfnamefont {C.~T.}\ \bibnamefont {Russell}}, \bibinfo {author}
  {\bibfnamefont {R.~J.}\ \bibnamefont {Strangeway}}, \bibinfo {author}
  {\bibfnamefont {R.~B.}\ \bibnamefont {Torbert}}, \bibinfo {author}
  {\bibfnamefont {B.~L.}\ \bibnamefont {Giles}}, \bibinfo {author}
  {\bibfnamefont {D.~J.}\ \bibnamefont {Gershman}}, \bibinfo {author}
  {\bibfnamefont {C.~J.}\ \bibnamefont {Pollock}}, \bibinfo {author}
  {\bibfnamefont {T.~E.}\ \bibnamefont {Moore}},\ and\ \bibinfo {author}
  {\bibfnamefont {J.~L.}\ \bibnamefont {Burch}},\ }\href
  {https://doi.org/10.3847/2041-8213/ab846e} {\bibfield  {journal} {\bibinfo
  {journal} {Astrophys.\ J. Lett.}\ }\textbf {\bibinfo {volume} {893}},\
  \bibinfo {pages} {L25} (\bibinfo {year} {2020})}\BibitemShut {NoStop}%
\bibitem [{\citenamefont {Huang}\ \emph {et~al.}(2020)\citenamefont {Huang},
  \citenamefont {Zhang}, \citenamefont {Sahraoui}, \citenamefont {Yuan},
  \citenamefont {Deng}, \citenamefont {Jiang}, \citenamefont {Xu},
  \citenamefont {Wei}, \citenamefont {He},\ and\ \citenamefont
  {Zhang}}]{HuangEA20-Bcurve}%
  \BibitemOpen
  \bibfield  {author} {\bibinfo {author} {\bibfnamefont {S.~Y.}\ \bibnamefont
  {Huang}}, \bibinfo {author} {\bibfnamefont {J.}~\bibnamefont {Zhang}},
  \bibinfo {author} {\bibfnamefont {F.}~\bibnamefont {Sahraoui}}, \bibinfo
  {author} {\bibfnamefont {Z.~G.}\ \bibnamefont {Yuan}}, \bibinfo {author}
  {\bibfnamefont {X.~H.}\ \bibnamefont {Deng}}, \bibinfo {author}
  {\bibfnamefont {K.}~\bibnamefont {Jiang}}, \bibinfo {author} {\bibfnamefont
  {S.~B.}\ \bibnamefont {Xu}}, \bibinfo {author} {\bibfnamefont {Y.~Y.}\
  \bibnamefont {Wei}}, \bibinfo {author} {\bibfnamefont {L.~H.}\ \bibnamefont
  {He}},\ and\ \bibinfo {author} {\bibfnamefont {Z.~H.}\ \bibnamefont
  {Zhang}},\ }\href {https://doi.org/10.3847/2041-8213/aba263} {\bibfield
  {journal} {\bibinfo  {journal} {Astrophys.\ J. Lett.}\ }\textbf {\bibinfo
  {volume} {898}},\ \bibinfo {pages} {L18} (\bibinfo {year}
  {2020})}\BibitemShut {NoStop}%
\bibitem [{\citenamefont {Ji}\ \emph {et~al.}(2022)\citenamefont {Ji},
  \citenamefont {Shen}, \citenamefont {Ren}, \citenamefont {Ma}, \citenamefont
  {Ma},\ and\ \citenamefont {Chen}}]{Ji2022}%
  \BibitemOpen
  \bibfield  {author} {\bibinfo {author} {\bibfnamefont {Y.}~\bibnamefont
  {Ji}}, \bibinfo {author} {\bibfnamefont {C.}~\bibnamefont {Shen}}, \bibinfo
  {author} {\bibfnamefont {N.}~\bibnamefont {Ren}}, \bibinfo {author}
  {\bibfnamefont {L.}~\bibnamefont {Ma}}, \bibinfo {author} {\bibfnamefont
  {Y.~H.}\ \bibnamefont {Ma}},\ and\ \bibinfo {author} {\bibfnamefont
  {X.}~\bibnamefont {Chen}},\ }\href@noop {} {\bibfield  {journal} {\bibinfo
  {journal} {Astrophys.\ J.}\ ,\ \bibinfo {pages} {67}} (\bibinfo {year}
  {2022})}\BibitemShut {NoStop}%
\bibitem [{\citenamefont {Hengster}(2024)}]{Hengster2024Thesis}%
  \BibitemOpen
  \bibfield  {author} {\bibinfo {author} {\bibfnamefont {Y.}~\bibnamefont
  {Hengster}},\ }\emph {\bibinfo {title} {Curvature Statistics of
  Turbulence}},\ \href@noop {} {Ph.D. thesis},\ \bibinfo  {school} {School of
  Mathematics, The University of Edinburgh} (\bibinfo {year}
  {2024})\BibitemShut {NoStop}%
\bibitem [{\citenamefont {Bentkamp}\ \emph {et~al.}(2022)\citenamefont
  {Bentkamp}, \citenamefont {Drivas}, \citenamefont {Lalescu},\ and\
  \citenamefont {Wilczek}}]{Bentkamp2022}%
  \BibitemOpen
  \bibfield  {author} {\bibinfo {author} {\bibfnamefont {L.}~\bibnamefont
  {Bentkamp}}, \bibinfo {author} {\bibfnamefont {T.}~\bibnamefont {Drivas}},
  \bibinfo {author} {\bibfnamefont {C.}~\bibnamefont {Lalescu}},\ and\ \bibinfo
  {author} {\bibfnamefont {M.}~\bibnamefont {Wilczek}},\ }\href@noop {}
  {\bibfield  {journal} {\bibinfo  {journal} {Nat. Commun.}\ }\textbf {\bibinfo
  {volume} {13}},\ \bibinfo {pages} {2088} (\bibinfo {year}
  {2022})}\BibitemShut {NoStop}%
\bibitem [{\citenamefont {Bentkamp}\ \emph {et~al.}(2019)\citenamefont
  {Bentkamp}, \citenamefont {Lalescu},\ and\ \citenamefont
  {Wilczek}}]{Bentkamp2019}%
  \BibitemOpen
  \bibfield  {author} {\bibinfo {author} {\bibfnamefont {L.}~\bibnamefont
  {Bentkamp}}, \bibinfo {author} {\bibfnamefont {C.~C.}\ \bibnamefont
  {Lalescu}},\ and\ \bibinfo {author} {\bibfnamefont {M.}~\bibnamefont
  {Wilczek}},\ }\href@noop {} {\bibfield  {journal} {\bibinfo  {journal} {Nat.
  Commun.}\ }\textbf {\bibinfo {volume} {10}},\ \bibinfo {pages} {3550}
  (\bibinfo {year} {2019})}\BibitemShut {NoStop}%
\bibitem [{\citenamefont {Godbersen}\ \emph {et~al.}(2021)\citenamefont
  {Godbersen}, \citenamefont {Bosbach}, \citenamefont {Schanz},\ and\
  \citenamefont {Schr\"oder}}]{Godbersen2021}%
  \BibitemOpen
  \bibfield  {author} {\bibinfo {author} {\bibfnamefont {P.}~\bibnamefont
  {Godbersen}}, \bibinfo {author} {\bibfnamefont {J.}~\bibnamefont {Bosbach}},
  \bibinfo {author} {\bibfnamefont {D.}~\bibnamefont {Schanz}},\ and\ \bibinfo
  {author} {\bibfnamefont {A.}~\bibnamefont {Schr\"oder}},\ }\href@noop {}
  {\bibfield  {journal} {\bibinfo  {journal} {Phys. Rev. Fluids}\ }\textbf
  {\bibinfo {volume} {6}},\ \bibinfo {pages} {110509} (\bibinfo {year}
  {2021})}\BibitemShut {NoStop}%
\bibitem [{\citenamefont {La~Porta}\ \emph {et~al.}(2001)\citenamefont
  {La~Porta}, \citenamefont {Voth}, \citenamefont {Crawford}, \citenamefont
  {Alexander},\ and\ \citenamefont {Bodenschatz}}]{LaPorta2001}%
  \BibitemOpen
  \bibfield  {author} {\bibinfo {author} {\bibfnamefont {A.}~\bibnamefont
  {La~Porta}}, \bibinfo {author} {\bibfnamefont {G.~A.}\ \bibnamefont {Voth}},
  \bibinfo {author} {\bibfnamefont {A.~M.}\ \bibnamefont {Crawford}}, \bibinfo
  {author} {\bibfnamefont {J.}~\bibnamefont {Alexander}},\ and\ \bibinfo
  {author} {\bibfnamefont {E.}~\bibnamefont {Bodenschatz}},\ }\href@noop {}
  {\bibfield  {journal} {\bibinfo  {journal} {Nature}\ }\textbf {\bibinfo
  {volume} {409}},\ \bibinfo {pages} {1017} (\bibinfo {year}
  {2001})}\BibitemShut {NoStop}%
\bibitem [{\citenamefont {Voth}\ \emph {et~al.}(2002)\citenamefont {Voth},
  \citenamefont {La~Porta}, \citenamefont {Crawford}, \citenamefont
  {Bodenschatz},\ and\ \citenamefont {Alexander}}]{Voth2002}%
  \BibitemOpen
  \bibfield  {author} {\bibinfo {author} {\bibfnamefont {G.~A.}\ \bibnamefont
  {Voth}}, \bibinfo {author} {\bibfnamefont {A.}~\bibnamefont {La~Porta}},
  \bibinfo {author} {\bibfnamefont {A.}~\bibnamefont {Crawford}}, \bibinfo
  {author} {\bibfnamefont {E.}~\bibnamefont {Bodenschatz}},\ and\ \bibinfo
  {author} {\bibfnamefont {J.}~\bibnamefont {Alexander}},\ }\href@noop {}
  {\bibfield  {journal} {\bibinfo  {journal} {J.\ Fluid Mech.}\ }\textbf
  {\bibinfo {volume} {469}},\ \bibinfo {pages} {121–160} (\bibinfo {year}
  {2002})}\BibitemShut {NoStop}%
\bibitem [{\citenamefont {Mordant}\ \emph {et~al.}(2004)\citenamefont
  {Mordant}, \citenamefont {Crawford},\ and\ \citenamefont
  {Bodenschatz}}]{Mordant2004a}%
  \BibitemOpen
  \bibfield  {author} {\bibinfo {author} {\bibfnamefont {N.}~\bibnamefont
  {Mordant}}, \bibinfo {author} {\bibfnamefont {A.}~\bibnamefont {Crawford}},\
  and\ \bibinfo {author} {\bibfnamefont {E.}~\bibnamefont {Bodenschatz}},\
  }\href@noop {} {\bibfield  {journal} {\bibinfo  {journal} {Phys.\ Rev.\
  Lett.}\ }\textbf {\bibinfo {volume} {93}},\ \bibinfo {pages} {214501}
  (\bibinfo {year} {2004})}\BibitemShut {NoStop}%
\bibitem [{\citenamefont {Homann}\ \emph {et~al.}(2011)\citenamefont {Homann},
  \citenamefont {Schulz},\ and\ \citenamefont {Grauer}}]{Homann2011}%
  \BibitemOpen
  \bibfield  {author} {\bibinfo {author} {\bibfnamefont {H.}~\bibnamefont
  {Homann}}, \bibinfo {author} {\bibfnamefont {D.}~\bibnamefont {Schulz}},\
  and\ \bibinfo {author} {\bibfnamefont {R.}~\bibnamefont {Grauer}},\
  }\href@noop {} {\bibfield  {journal} {\bibinfo  {journal} {Phys.\ Fluids}\
  }\textbf {\bibinfo {volume} {23}},\ \bibinfo {pages} {055102} (\bibinfo
  {year} {2011})}\BibitemShut {NoStop}%
\bibitem [{\citenamefont {Crawford}\ \emph {et~al.}(2005)\citenamefont
  {Crawford}, \citenamefont {Mordant},\ and\ \citenamefont
  {Bodenschatz}}]{Crawford2005}%
  \BibitemOpen
  \bibfield  {author} {\bibinfo {author} {\bibfnamefont {A.~M.}\ \bibnamefont
  {Crawford}}, \bibinfo {author} {\bibfnamefont {N.}~\bibnamefont {Mordant}},\
  and\ \bibinfo {author} {\bibfnamefont {E.}~\bibnamefont {Bodenschatz}},\
  }\href@noop {} {\bibfield  {journal} {\bibinfo  {journal} {Phys.\ Rev.\
  Lett.}\ }\textbf {\bibinfo {volume} {94}},\ \bibinfo {pages} {024501}
  (\bibinfo {year} {2005})}\BibitemShut {NoStop}%
\bibitem [{\citenamefont {Gagne}\ \emph {et~al.}(1994)\citenamefont {Gagne},
  \citenamefont {Marchand},\ and\ \citenamefont {Castaing}}]{Gagne1994}%
  \BibitemOpen
  \bibfield  {author} {\bibinfo {author} {\bibfnamefont {Y.}~\bibnamefont
  {Gagne}}, \bibinfo {author} {\bibfnamefont {M.}~\bibnamefont {Marchand}},\
  and\ \bibinfo {author} {\bibfnamefont {B.}~\bibnamefont {Castaing}},\
  }\href@noop {} {\bibfield  {journal} {\bibinfo  {journal} {J. Phys. II Fr.}\
  }\textbf {\bibinfo {volume} {4}},\ \bibinfo {pages} {1} (\bibinfo {year}
  {1994})}\BibitemShut {NoStop}%
\bibitem [{\citenamefont {Naert}\ \emph {et~al.}(1998)\citenamefont {Naert},
  \citenamefont {Castaing}, \citenamefont {Chabaud}, \citenamefont
  {H{\'e}bral},\ and\ \citenamefont {Peinke}}]{Naert1998}%
  \BibitemOpen
  \bibfield  {author} {\bibinfo {author} {\bibfnamefont {A.}~\bibnamefont
  {Naert}}, \bibinfo {author} {\bibfnamefont {B.}~\bibnamefont {Castaing}},
  \bibinfo {author} {\bibfnamefont {B.}~\bibnamefont {Chabaud}}, \bibinfo
  {author} {\bibfnamefont {B.}~\bibnamefont {H{\'e}bral}},\ and\ \bibinfo
  {author} {\bibfnamefont {J.}~\bibnamefont {Peinke}},\ }\href@noop {}
  {\bibfield  {journal} {\bibinfo  {journal} {Physica D}\ }\textbf {\bibinfo
  {volume} {113}},\ \bibinfo {pages} {73} (\bibinfo {year} {1998})}\BibitemShut
  {NoStop}%
\bibitem [{\citenamefont {Lawson}\ \emph {et~al.}(2019)\citenamefont {Lawson},
  \citenamefont {Bodenschatz}, \citenamefont {Knutsen}, \citenamefont
  {Dawson},\ and\ \citenamefont {Worth}}]{Lawson2019}%
  \BibitemOpen
  \bibfield  {author} {\bibinfo {author} {\bibfnamefont {J.~M.}\ \bibnamefont
  {Lawson}}, \bibinfo {author} {\bibfnamefont {E.}~\bibnamefont {Bodenschatz}},
  \bibinfo {author} {\bibfnamefont {A.~N.}\ \bibnamefont {Knutsen}}, \bibinfo
  {author} {\bibfnamefont {J.~R.}\ \bibnamefont {Dawson}},\ and\ \bibinfo
  {author} {\bibfnamefont {N.~A.}\ \bibnamefont {Worth}},\ }\href@noop {}
  {\bibfield  {journal} {\bibinfo  {journal} {Phys.\ Rev.\ Fluids}\ }\textbf
  {\bibinfo {volume} {4}},\ \bibinfo {pages} {022601} (\bibinfo {year}
  {2019})}\BibitemShut {NoStop}%
\bibitem [{\citenamefont {Kolmogorov}(1962)}]{Kolmogorov62}%
  \BibitemOpen
  \bibfield  {author} {\bibinfo {author} {\bibfnamefont {A.~N.}\ \bibnamefont
  {Kolmogorov}},\ }\href@noop {} {\bibfield  {journal} {\bibinfo  {journal}
  {J.\ Fluid Mech.}\ }\textbf {\bibinfo {volume} {13}},\ \bibinfo {pages} {82}
  (\bibinfo {year} {1962})}\BibitemShut {NoStop}%
\bibitem [{\citenamefont {Oboukhov}(1962)}]{Oboukhov62}%
  \BibitemOpen
  \bibfield  {author} {\bibinfo {author} {\bibfnamefont {A.~M.}\ \bibnamefont
  {Oboukhov}},\ }\href {https://doi.org/10.1017/S0022112062000506} {\bibfield
  {journal} {\bibinfo  {journal} {J.\ Fluid Mech.}\ }\textbf {\bibinfo {volume}
  {13}},\ \bibinfo {pages} {77–81} (\bibinfo {year} {1962})}\BibitemShut
  {NoStop}%
\bibitem [{\citenamefont {Schr{\"o}der}\ \emph {et~al.}(2022)\citenamefont
  {Schr{\"o}der}, \citenamefont {Schanz}, \citenamefont {Gesemann},
  \citenamefont {Huhn}, \citenamefont {Buchwald}, \citenamefont {Paz},\ and\
  \citenamefont {Bodenschatz}}]{Schroeder2022}%
  \BibitemOpen
  \bibfield  {author} {\bibinfo {author} {\bibfnamefont {A.}~\bibnamefont
  {Schr{\"o}der}}, \bibinfo {author} {\bibfnamefont {D.}~\bibnamefont
  {Schanz}}, \bibinfo {author} {\bibfnamefont {S.}~\bibnamefont {Gesemann}},
  \bibinfo {author} {\bibfnamefont {F.}~\bibnamefont {Huhn}}, \bibinfo {author}
  {\bibfnamefont {T.}~\bibnamefont {Buchwald}}, \bibinfo {author}
  {\bibfnamefont {D.~G.}\ \bibnamefont {Paz}},\ and\ \bibinfo {author}
  {\bibfnamefont {E.}~\bibnamefont {Bodenschatz}},\ }\href@noop {} {\bibfield
  {journal} {\bibinfo  {journal} {Proceedings of 20th International Symposium
  on Application of Laser and Imaging Techniques to Fluid Mechanics}\ }
  (\bibinfo {year} {2022})},\ \bibinfo {note} {lisbon, Portugal}\BibitemShut
  {NoStop}%
\bibitem [{\citenamefont {Bosbach}\ \emph {et~al.}(2021)\citenamefont
  {Bosbach}, \citenamefont {Schanz}, \citenamefont {Godbersen},\ and\
  \citenamefont {Schr{\"o}der}}]{Bosbach2021}%
  \BibitemOpen
  \bibfield  {author} {\bibinfo {author} {\bibfnamefont {J.}~\bibnamefont
  {Bosbach}}, \bibinfo {author} {\bibfnamefont {D.}~\bibnamefont {Schanz}},
  \bibinfo {author} {\bibfnamefont {P.}~\bibnamefont {Godbersen}},\ and\
  \bibinfo {author} {\bibfnamefont {A.}~\bibnamefont {Schr{\"o}der}},\
  }\href@noop {} {\bibfield  {journal} {\bibinfo  {journal} {Proceedings of
  14th International Symposium on Particle Image Velocimetry - ISPIV 2021}\ }
  (\bibinfo {year} {2021})},\ \bibinfo {note} {{C}hicago, {USA}}\BibitemShut
  {NoStop}%
\bibitem [{\citenamefont {Weiss}\ \emph {et~al.}(2024)\citenamefont {Weiss},
  \citenamefont {Schanz}, \citenamefont {Erdogdu}, \citenamefont {Schr\"oder},\
  and\ \citenamefont {Bosbach}}]{Weiss2024}%
  \BibitemOpen
  \bibfield  {author} {\bibinfo {author} {\bibfnamefont {S.}~\bibnamefont
  {Weiss}}, \bibinfo {author} {\bibfnamefont {D.}~\bibnamefont {Schanz}},
  \bibinfo {author} {\bibfnamefont {A.~O.}\ \bibnamefont {Erdogdu}}, \bibinfo
  {author} {\bibfnamefont {A.}~\bibnamefont {Schr\"oder}},\ and\ \bibinfo
  {author} {\bibfnamefont {J.}~\bibnamefont {Bosbach}},\ }\href
  {https://doi.org/10.1017/jfm.2024.677} {\bibfield  {journal} {\bibinfo
  {journal} {J. Fluid Mech.}\ }\textbf {\bibinfo {volume} {999}},\ \bibinfo
  {pages} {A90} (\bibinfo {year} {2024})}\BibitemShut {NoStop}%
\bibitem [{\citenamefont {{Schanz}}\ \emph {et~al.}(2016)\citenamefont
  {{Schanz}}, \citenamefont {{Gesemann}},\ and\ \citenamefont
  {{Schr{\"o}der}}}]{Schanz2016}%
  \BibitemOpen
  \bibfield  {author} {\bibinfo {author} {\bibfnamefont {D.}~\bibnamefont
  {{Schanz}}}, \bibinfo {author} {\bibfnamefont {S.}~\bibnamefont
  {{Gesemann}}},\ and\ \bibinfo {author} {\bibfnamefont {A.}~\bibnamefont
  {{Schr{\"o}der}}},\ }\href@noop {} {\bibfield  {journal} {\bibinfo  {journal}
  {Exp. Fluids}\ }\textbf {\bibinfo {volume} {57}},\ \bibinfo {pages} {70}
  (\bibinfo {year} {2016})}\BibitemShut {NoStop}%
\bibitem [{\citenamefont {Schröder}\ and\ \citenamefont
  {Schanz}(2023)}]{Schroeder2023}%
  \BibitemOpen
  \bibfield  {author} {\bibinfo {author} {\bibfnamefont {A.}~\bibnamefont
  {Schröder}}\ and\ \bibinfo {author} {\bibfnamefont {D.}~\bibnamefont
  {Schanz}},\ }\href@noop {} {\bibfield  {journal} {\bibinfo  {journal} {Annu.
  Rev. Fluid Mech.}\ }\textbf {\bibinfo {volume} {55}},\ \bibinfo {pages} {511}
  (\bibinfo {year} {2023})}\BibitemShut {NoStop}%
\bibitem [{\citenamefont {Wieneke}(2012)}]{Wieneke2013}%
  \BibitemOpen
  \bibfield  {author} {\bibinfo {author} {\bibfnamefont {B.}~\bibnamefont
  {Wieneke}},\ }\href@noop {} {\bibfield  {journal} {\bibinfo  {journal} {Meas.
  Sci. Technol}\ }\textbf {\bibinfo {volume} {24}},\ \bibinfo {pages} {024008}
  (\bibinfo {year} {2012})}\BibitemShut {NoStop}%
\bibitem [{\citenamefont {{Jahn}}\ \emph {et~al.}(2021)\citenamefont {{Jahn}},
  \citenamefont {{Schanz}},\ and\ \citenamefont {{Schr{\"o}der}}}]{Jahn2021}%
  \BibitemOpen
  \bibfield  {author} {\bibinfo {author} {\bibfnamefont {T.}~\bibnamefont
  {{Jahn}}}, \bibinfo {author} {\bibfnamefont {D.}~\bibnamefont {{Schanz}}},\
  and\ \bibinfo {author} {\bibfnamefont {A.}~\bibnamefont {{Schr{\"o}der}}},\
  }\href@noop {} {\bibfield  {journal} {\bibinfo  {journal} {Exp. Fluids}\
  }\textbf {\bibinfo {volume} {62}},\ \bibinfo {pages} {179} (\bibinfo {year}
  {2021})}\BibitemShut {NoStop}%
\bibitem [{\citenamefont {Gesemann}\ \emph {et~al.}(2016)\citenamefont
  {Gesemann}, \citenamefont {Huhn}, \citenamefont {Schanz},\ and\ \citenamefont
  {Schröder}}]{Gesemann2016}%
  \BibitemOpen
  \bibfield  {author} {\bibinfo {author} {\bibfnamefont {S.}~\bibnamefont
  {Gesemann}}, \bibinfo {author} {\bibfnamefont {F.}~\bibnamefont {Huhn}},
  \bibinfo {author} {\bibfnamefont {D.}~\bibnamefont {Schanz}},\ and\ \bibinfo
  {author} {\bibfnamefont {A.}~\bibnamefont {Schröder}},\ }\href@noop {}
  {\bibfield  {journal} {\bibinfo  {journal} {{Proceedings of 18th
  International Symposium on Applications of Laser Techniques to Fluid
  Mechanics}}\ } (\bibinfo {year} {2016})},\ \bibinfo {note} {lisbon,
  Portugal}\BibitemShut {NoStop}%
\bibitem [{\citenamefont {{Ouellette}}\ \emph {et~al.}(2006)\citenamefont
  {{Ouellette}}, \citenamefont {{Xu}}, \citenamefont {{Bourgoin}},\ and\
  \citenamefont {{Bodenschatz}}}]{Ouellette2006}%
  \BibitemOpen
  \bibfield  {author} {\bibinfo {author} {\bibfnamefont {N.~T.}\ \bibnamefont
  {{Ouellette}}}, \bibinfo {author} {\bibfnamefont {H.}~\bibnamefont {{Xu}}},
  \bibinfo {author} {\bibfnamefont {M.}~\bibnamefont {{Bourgoin}}},\ and\
  \bibinfo {author} {\bibfnamefont {E.}~\bibnamefont {{Bodenschatz}}},\
  }\href@noop {} {\bibfield  {journal} {\bibinfo  {journal} {New J.\ Phys.}\
  }\textbf {\bibinfo {volume} {8}},\ \bibinfo {pages} {102} (\bibinfo {year}
  {2006})}\BibitemShut {NoStop}%
\bibitem [{\citenamefont {Kallenberg}(2021)}]{Kallenberg}%
  \BibitemOpen
  \bibfield  {author} {\bibinfo {author} {\bibfnamefont {O.}~\bibnamefont
  {Kallenberg}},\ }\href@noop {} {\emph {\bibinfo {title} {Foundations of
  Modern Probability}}},\ \bibinfo {edition} {3rd}\ ed.\ (\bibinfo  {publisher}
  {Springer Cham},\ \bibinfo {year} {2021})\BibitemShut {NoStop}%
\bibitem [{\citenamefont {Bos}\ \emph {et~al.}(2015)\citenamefont {Bos},
  \citenamefont {Kadoch},\ and\ \citenamefont {Schneider}}]{Bos2015}%
  \BibitemOpen
  \bibfield  {author} {\bibinfo {author} {\bibfnamefont {W.}~\bibnamefont
  {Bos}}, \bibinfo {author} {\bibfnamefont {B.}~\bibnamefont {Kadoch}},\ and\
  \bibinfo {author} {\bibfnamefont {K.}~\bibnamefont {Schneider}},\ }\href@noop
  {} {\bibfield  {journal} {\bibinfo  {journal} {Phys.~Rev.~Lett.}\ }\textbf
  {\bibinfo {volume} {114}},\ \bibinfo {pages} {214502} (\bibinfo {year}
  {2015})}\BibitemShut {NoStop}%
\bibitem [{\citenamefont {L\"ubke}\ \emph {et~al.}(2024)\citenamefont
  {L\"ubke}, \citenamefont {Effenberger}, \citenamefont {Wilbert},
  \citenamefont {Fichtner},\ and\ \citenamefont {Grauer}}]{Luebke2024}%
  \BibitemOpen
  \bibfield  {author} {\bibinfo {author} {\bibfnamefont {J.}~\bibnamefont
  {L\"ubke}}, \bibinfo {author} {\bibfnamefont {F.}~\bibnamefont
  {Effenberger}}, \bibinfo {author} {\bibfnamefont {M.}~\bibnamefont
  {Wilbert}}, \bibinfo {author} {\bibfnamefont {H.}~\bibnamefont {Fichtner}},\
  and\ \bibinfo {author} {\bibfnamefont {R.}~\bibnamefont {Grauer}},\ }\href
  {https://doi.org/10.1209/0295-5075/ad438f} {\bibfield  {journal} {\bibinfo
  {journal} {Europhys. Lett.}\ }\textbf {\bibinfo {volume} {146}},\ \bibinfo
  {pages} {43001} (\bibinfo {year} {2024})}\BibitemShut {NoStop}%
\bibitem [{\citenamefont {Kadoch}\ \emph {et~al.}(2022)\citenamefont {Kadoch},
  \citenamefont {del Castillo-Negrete}, \citenamefont {Bos},\ and\
  \citenamefont {Schneider}}]{Kadoch_PoP_2022}%
  \BibitemOpen
  \bibfield  {author} {\bibinfo {author} {\bibfnamefont {B.}~\bibnamefont
  {Kadoch}}, \bibinfo {author} {\bibfnamefont {D.}~\bibnamefont {del
  Castillo-Negrete}}, \bibinfo {author} {\bibfnamefont {W.~J.~T.}\ \bibnamefont
  {Bos}},\ and\ \bibinfo {author} {\bibfnamefont {K.}~\bibnamefont
  {Schneider}},\ }\href@noop {} {\bibfield  {journal} {\bibinfo  {journal}
  {Phys. Plasmas}\ }\textbf {\bibinfo {volume} {29}},\ \bibinfo {pages}
  {102301} (\bibinfo {year} {2022})}\BibitemShut {NoStop}%
\bibitem [{\citenamefont {Gheorghiu}\ \emph {et~al.}(2024)\citenamefont
  {Gheorghiu}, \citenamefont {Militello},\ and\ \citenamefont {{Juul
  Rasmussen}}}]{Gheorghiu_2024}%
  \BibitemOpen
  \bibfield  {author} {\bibinfo {author} {\bibfnamefont {T.}~\bibnamefont
  {Gheorghiu}}, \bibinfo {author} {\bibfnamefont {F.}~\bibnamefont
  {Militello}},\ and\ \bibinfo {author} {\bibfnamefont {J.}~\bibnamefont {{Juul
  Rasmussen}}},\ }\href@noop {} {\bibfield  {journal} {\bibinfo  {journal}
  {Phys Plasmas}\ }\textbf {\bibinfo {volume} {31}},\ \bibinfo {pages} {013901}
  (\bibinfo {year} {2024})}\BibitemShut {NoStop}%
\bibitem [{\citenamefont {Lin}\ \emph {et~al.}(2025)\citenamefont {Lin},
  \citenamefont {Kadoch}, \citenamefont {Benkadda},\ and\ \citenamefont
  {Schneider}}]{Lin_PPCF_2025}%
  \BibitemOpen
  \bibfield  {author} {\bibinfo {author} {\bibfnamefont {Z.}~\bibnamefont
  {Lin}}, \bibinfo {author} {\bibfnamefont {B.}~\bibnamefont {Kadoch}},
  \bibinfo {author} {\bibfnamefont {S.}~\bibnamefont {Benkadda}},\ and\
  \bibinfo {author} {\bibfnamefont {K.}~\bibnamefont {Schneider}},\ }\href@noop
  {} {\bibfield  {journal} {\bibinfo  {journal} {Plasma Phys. Control. Fusion}\
  }\textbf {\bibinfo {volume} {67}},\ \bibinfo {pages} {045038} (\bibinfo
  {year} {2025})}\BibitemShut {NoStop}%
\end{thebibliography}%

\end{document}